\documentclass[journal]{IEEEtran}

\usepackage{cite}
\usepackage{hyperref}

\ifCLASSINFOpdf
    \usepackage[pdftex]{graphicx}
    \graphicspath{{./figures/}}
\else
\fi

\usepackage{mathtools}
\usepackage{algorithm}
\usepackage{algorithmic}
\usepackage{eqparbox}

\usepackage{xcolor}
\usepackage{xspace} 

\ifCLASSOPTIONcompsoc
    \usepackage[caption=false,font=normalsize,labelfont=sf,textfont=sf]{subfig}
\else
  \usepackage[caption=false,font=footnotesize]{subfig}
\fi
\usepackage{siunitx}
\usepackage{mathptmx}
\usepackage{threeparttable}
\usepackage{amsmath,amssymb,amsfonts}
\usepackage{algorithmic}
\usepackage{textcomp}
\usepackage{array}
\usepackage{multirow}

\begin{document}
\bstctlcite{IEEEexample:BSTcontrol} 

\title{An Efficient Power Management Unit with Continuous MPPT and Energy Recycling for Wireless Millimetric Biomedical Implants}

\author{
        Yiwei~Zou,~\IEEEmembership{Student Member,~IEEE,}
        Huan-Cheng~Liao,~\IEEEmembership{Student Member,~IEEE,}
        Wei~Wang,~\IEEEmembership{Student Member,~IEEE,}
        Wonjune~Kim,~\IEEEmembership{Student Member,~IEEE,}
        Yumin~Su,~\IEEEmembership{Student Member,~IEEE,}
        Jacob~T.~Robinson,~\IEEEmembership{Senior~Member,~IEEE,}
        and~Kaiyuan~Yang,~\IEEEmembership{Senior~Member,~IEEE}
\thanks{Manuscript received on}
\thanks{All authors are with the Department of Electrical and Computer Engineering, Rice University, Houston, TX 77005, USA. (Corresponding author: Kaiyuan Yang, kyang@rice.edu)
}
\thanks{This work is supported in part by the National Science Foundation (NSF) awards 2023849 and 2146476. }
}

\markboth{Journal of Solid-State Circuits}%
{Shell \MakeLowercase{\textit{et al.}}: A Sample Article Using IEEEtran.cls for IEEE Journals}


\maketitle

\begin{abstract}
Biomedical implants offer transformative tools to improve medical outcomes. To realize minimally invasive implants with miniaturized volume and weight, wireless power transfer (WPT) has been extensively studied to replace bulky batteries that dominate the volume of traditional implants and require surgical replacements. Ultrasonic (US) and magnetoelectric (ME) WPT modalities, which leverage low-frequency acoustic-electrical coupling for energy transduction, become viable solutions for millimeter-scale receivers.
This work presents a fully integrated power management unit (PMU) for ME WPT in millimetric implants. The PMU achieves load-independent maximum power extraction and usage by continuously matching the transducer’s impedance, dynamically optimizing the power stage across varying input/load conditions, and reusing the storage energy to sustain the system when input power drops. Its parallel-input regulation and storing stages architecture prevents the cascading power loss. With the skewed-duty-cycle MPPT technique and regulation efficiency optimizer, the PMU achieves a peak MPPT efficiency of 98.5\% and a peak system overall efficiency of 73.33\%. Additionally, the PMU includes an adaptive high-voltage charging stage that charges the stimulation capacitor up to 12~V with an improved efficiency of 37.88\%.
\end{abstract}


\section{Introduction}

\IEEEPARstart{B}{iomedical} implants offer revolutionized healthcare by enabling long-term monitoring, diagnosis, and treatment of numerous challenging physiological conditions such as cardiovascular diseases \cite{kwon_battery-less_2023}, pain relief \cite{verrills_review_2016}, and neurological disorders \cite{lozano_deep_2019}. These chronically implanted devices can continuously collect physiological data and provide timely therapeutic interventions. However, powering these devices remains a significant challenge, particularly in the pursuit of minimally invasive implants that are smaller than a few millimeters in all dimensions. 
Conventional bio-implants have relied on batteries as their primary energy source. These bulky batteries limit the miniaturization of implant volumes and often necessitate surgical replacement once depleted. These issues have driven the research and development of wireless power transfer (WPT) technologies for battery-free miniature implants \cite{nair_miniature_2023}.

Inductive coupling, a prevalent WPT method for electronics, is well studied and benefits from advanced power management techniques, such as resonant regulating rectifiers \cite{lee_273_2024,kim_144-mhz_2017}, to enhance efficiency. However, it suffers from either large receiver (RX) coil sizes or high tissue absorption at higher carrier frequencies, which ultimately limits their applicability in deeply implanted miniature implants.

\begin{figure}[t!]
\centering
\includegraphics[width = 0.95\linewidth]{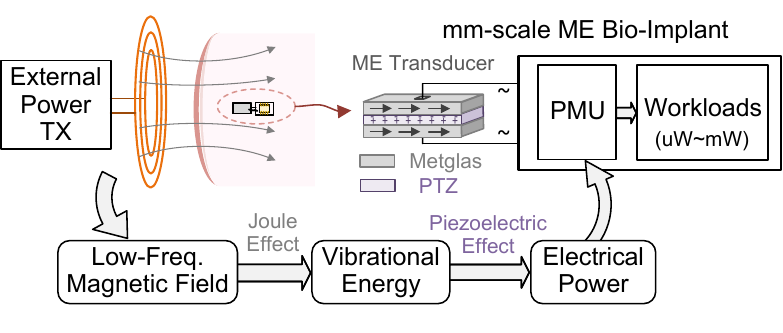}
\vskip -2ex
\caption{Magnetoelectric power transfer for mm-scale bio-implants \cite{yu_magni_2020, chen_wireless_2022}.}
\label{ME_WPT}
\end{figure}

Ultrasonic (US) \cite{seo_wireless_2016,piech_wireless_2020,sonmezoglu_45mm3_2020} and magnetoelectric (ME) \cite{yu_magni_2020, chen_wireless_2022} WPT modalities have emerged as viable solutions for mm-scale implants. They lower acoustic resonance frequencies (roughly between 0.1 to 10 MHz) and wavelengths for energy transduction, thus enjoying smaller receivers without tissue absorption. 
Compared to US WPT, the ME mechanism, as illustrated in Fig.~\ref{ME_WPT}, further avoids the reflection of ultrasounds at the boundaries between air, tissue, and bones \cite{hoskins2019diagnostic}.

Despite the promising characteristics of ME/US WPT modalities, power management units (PMUs) employed in recent demonstrations of battery-free wireless bio-implant systems have not been particularly optimized. Unlike conventional PMUs designed for battery-powered systems, PMUs for battery-less implants face a unique set of challenges. Firstly, owing to relatively low link efficiency, the power deliverable within safety limits is often restricted. Therefore, the PMU must convert power with the highest possible efficiency to supply loads typically in the milliwatt range. Moreover, the PMU must exhibit high adaptability to manage varying input conditions, stemming from wireless channel fluctuations, and dynamic output load demands. The PMU is also expected to be compatible with different types of transducers.
For implementation simplicity and operation robustness, recently demonstrated ME/US implants adopt basic rectifier-(DCDC)-LDO PMU topologies \cite{piech_wireless_2020,zhao_339_2024, yu_magnetoelectric_2022,yu_millimetric_2024}. However, this structure is inefficient when there is a mismatch between input power and output load. Thus, it is critical to develop specialized PMUs for millimeter-scale implants powered by ME/US WPT, taking into account the specific characteristics of the WPT method and the requirements of implantable devices. Specifically, we identify the following key requirements for the PMU design.


\begin{figure}[t!]
\centering
\subfloat[]{\includegraphics[width = 0.9\linewidth]{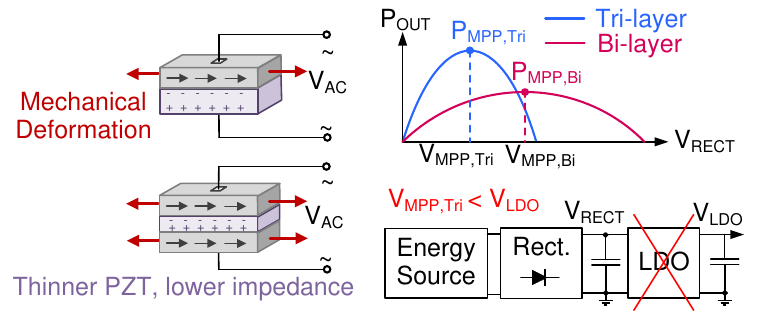}%
\label{transducer_codesign}}\\

\subfloat[]{\includegraphics[width = 0.9\linewidth]{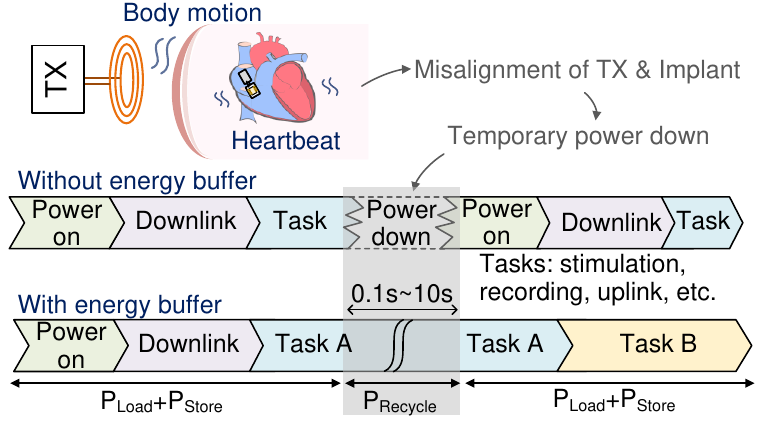}%
\label{body_motion}}\\

\subfloat[]{\includegraphics[width = 0.8\linewidth]{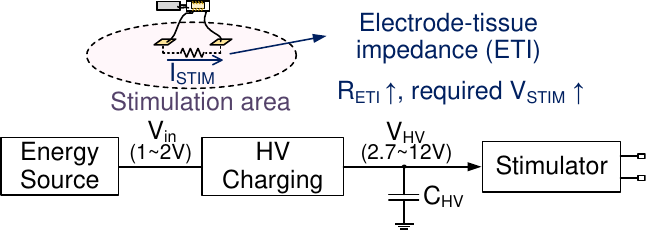}%
\label{eti_hv_charging}}\\

\caption{Demands of PMU design for ME/US WPT biomedical implants: (a) co-design of PMU and transducer; (b) miniature energy buffer for body motion tolerance; (c) efficient high voltage charging stage for neurostimulator.}
\label{PMU_demands}
\end{figure}

First, unlike inductive WPT, there exists an achievable maximum power point (MPP) in ME/US WPT, due to the transducers’ k$\Omega$ impedance. 
A recent study \cite{kim_magnetoelectrics_2023} reported an optimized tri-layer ME transducer that achieves higher power at MPP ($P_{MPP}$) and higher power transfer efficiency (PTE) than its bi-layer counterparts. The tri-layer film consists of a thinner lead zirconate titanate (PZT) layer sandwiched between two Metglas layers. The thin PZT layer increases the intrinsic capacitance $C_P$, resulting in a lower impedance, while the extra Metglas layer enhances magnetic-to-acoustic coupling and transduction efficiency.
However, this improvement comes at the cost of a lower output voltage. If using the traditional rectifier-LDO structure, the rectifier's output voltage might not be enough to provide enough headroom for the LDO, even if the power is sufficient. This suggests that the PMU for ME/US WPT should be co-designed with the transducer to achieve consistent operation at $P_{MPP}$ under varying input and load conditions. 

Second, WPT in miniature implants is sensitive to misalignment with the transmitter (TX) due to body motions (movement, heartbeat, respiration, etc.) \cite{wang_171_2024}. These events may cause temporary power drops and interruptions to the implant’s operation and require reprogramming of the system after the power is restored, shown in Fig.~\ref{body_motion}. Thus, an energy buffer should be considered in a WPT PMU for implants. Given the frequency range of human motions (typically 0.1$\sim$20~Hz)~\cite{khusainov_real-time_2013}, a miniature supercapacitor is an ideal choice for mm-size implants. The energy buffer also allows for the recycling of the excess power ($P_{MPP}$-$P_{load}$), which is wasted in traditional buffer-less PMUs, enhancing the system efficiency and extending the battery lifetime of the external TX. 

Thirdly, neuromodulation is a common task for bio-implants, where the required stimulation voltage is decided by the electrode-tissue impedance (ETI) conditions and the stimulation threshold, as shown in Fig.~\ref{eti_hv_charging}. For instance, epidural stimulation demands a stimulation voltage up to 14~V to activate the motor cortex, according to \cite{woods_miniature_2024}. Hence, a high-voltage (HV) charging stage in the PMU is desirable to efficiently charge a stimulation capacitor without damaging other components.

\begin{figure}[!t]
\centering
\includegraphics[width=\linewidth]{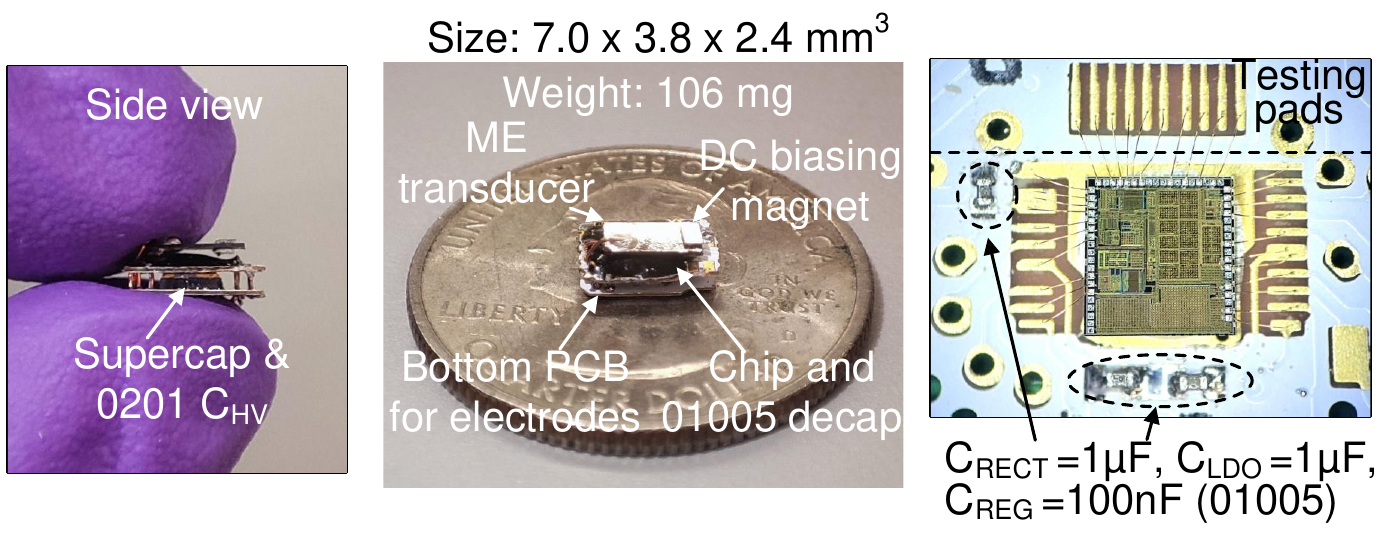}
\vskip -1ex
\caption{Prototyped mm-scale implant packaged with the $5\times 2\times 0.2~mm^3$ tri-layer ME transducer, supercapacitor, 0201 stimulation capacitor, and the three 01005 decoupling capacitors.}
\label{Package}
\end{figure}

To meet these demands, we present a fully integrated PMU tailored for wireless mm-scale bio-implants \cite{zou_parallel-input_2025}. It achieves load-independent maximum power extraction and usage by continuously matching the transducer’s impedance, dynamically optimizing the power stage's efficiency across varying input/load conditions, and reusing the storage energy to sustain the system when the input power drops. The proposed PMU offers five unique features: (1) an architecture with parallel-input regulation (REG) and storing (STO) stages to prevent cascading loss in series power stage topologies; (2) a capacitance redistribution topology maximizing the utilization of the on-chip capacitor; (3) a continuous skewed-duty-based MPPT (SD-MPPT) technique for ME transducers; (4) a real-time regulation efficiency ($\eta_{REG}$) optimizer that adjusts REG stage conversion ratio for optimal efficiency; and (5) an adaptive high-voltage charging stage that charges the stimulation capacitor up to 12~V with an adaptive conversion ratio for improved charging efficiency and speed. In addition to standalone PMU testing, we demonstrated a fully integrated ME-powered mm-scale implant capable of wireless communication, electrical stimulation, and recording, packaged in a $7\times 3.8\times 2.4~mm^3$ volume including all the passives, as shown in Fig.~\ref{Package}. While this work is prototyped for ME WPT, most of the presented analysis and techniques can be extended to US WPT bio-implants due to the similar impedance characteristics of ME and US transducers.

The rest of the article is organized as follows: Section II introduces the modeling and analysis of ME/US WPT, including the derivation of the maximum power point of the transducers. Section III elaborates on the system design and circuit implementation. Section IV presents the measurement results. Section V concludes the article.
\section{ME/US WPT Modeling and MPP Analysis}
\label{sec:analysis}

To ensure continuous operation at the optimal power point under varying environmental conditions and transducer characteristics, maximum power point tracking (MPPT) techniques are essential in PMU design. Traditional methods such as perturb and observe (P\&O) \cite{morel_322_2020,li_piezoelectric_2019,kim_energy-efficient_2013} and fractional open-circuit voltage (FOCV) \cite{shim_self-powered_2015,li_32na_2022,noh_reconfigurable_2022} either require complex power monitoring and digital control, or interrupt power delivery by temporarily disconnecting the transducer. More recently, a duty-cycle-based MPPT technique was proposed for piezoelectric harvesting (PEH), enabling continuous MPPT by regulating the rectification duty cycle to 50\% \cite{yue_303_2023}. However, its applicability to ME/US transducers remains uncertain due to their distinct impedance characteristics.

In addition, carrier frequency selection is critical in wireless power transfer systems. Unlike inductive coupling, where the resonance frequency is well defined by the LC components, ME/US transducers typically exhibit both short-circuit and open-circuit resonance frequencies. It remains unclear which frequency maximizes output power or voltage, complicating the design process.

These challenges motivate the development of a comprehensive analytical model that captures the behavior of both the transducer and its interface circuitry, providing a solid foundation for optimizing PMU performance.

\subsection{Modeling of ME/US WPT}

\begin{figure}[!t]
\centering
\includegraphics[width=0.8\linewidth]{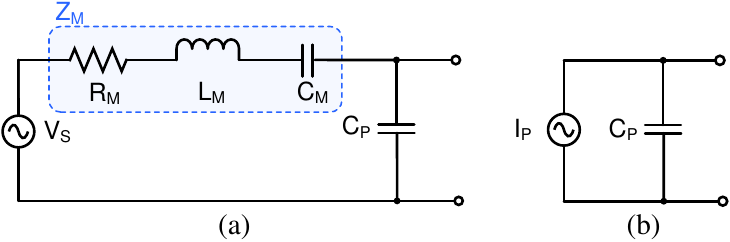}
\vskip -1ex
\caption{(a) Generic Butterworth Van-Dyke (BVD) model of ME/US transducers. (b) Simplified uncoupled model used in most PEH studies.}
\label{impedance_model}
\end{figure}

\begin{figure}[!t]
\centering
\includegraphics[width=0.9\linewidth]{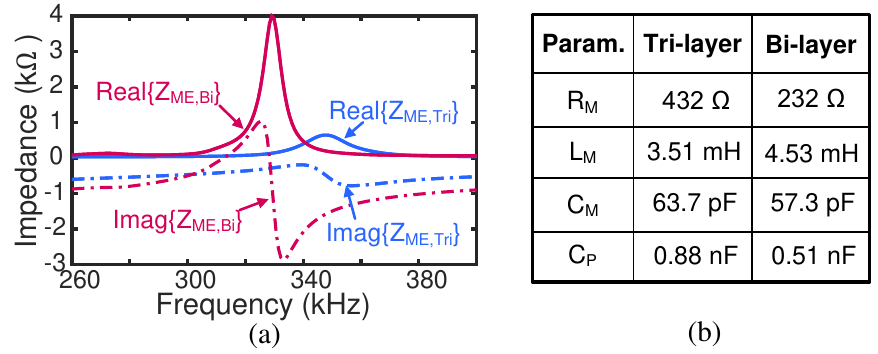}
\vskip -1ex
\caption{(a) Impedance measurement waveform of a tri-layer and a bi-layer ME transducer. (b) Model parameters obtained by fitting the impedance measurement.}
\label{impedance_measurement}
\end{figure}


Due to their acoustic-electrical coupling characteristics, ME/US transducers can be modeled using the Butterworth–Van Dyke (BVD) equivalent circuit \cite{noauthor_ieee_1978}, as shown in Fig.~\ref{impedance_model}a. In this model, the voltage source represents the transduction mechanism and is proportional to the energy input—linearly related to magnetic flux density for ME transducers and to acoustic pressure for ultrasound transducers. When analyzing the WPT link, this voltage source can be considered constant if the relative positioning between the transmitter and receiver remains fixed.
The remaining model parameters ($R_M$, $L_M$, $C_M$, $C_P$) can be extracted by measuring the transducer’s impedance and fitting both the real and imaginary components to the model \cite{yu_miniature_2024}. As an example, Fig.~\ref{impedance_measurement} shows the measured impedance profiles of tri-layer and bi-layer ME transducers along with the corresponding fitted BVD model parameters.


In piezoelectric energy harvesting, the transducer is often modeled using a simplified uncoupled model—a sinusoidal current source in parallel with a capacitor $C_P$, as shown in Fig.~\ref{impedance_model}b. This approximation is valid for piezoelectric harvesters with weak acoustic-electrical coupling, where changes in the electrical load have negligible effects on the transducer’s mechanical vibration. However, this assumption does not necessarily hold for ME or US transducers. The degree of coupling can be quantified by the following parameter \cite{shu_analysis_2006}: 
$\frac{k_{e}^{2}}{\zeta} = 2\sqrt{L_{M}C_{M}}/(C_{P}R_{M})$,
where $k_{e}^{2}$ is the electromechanical coupling coefficient and $\zeta$ is the damping ratio, both of which are determined by the material constant and structural geometry. If $\frac{k_{e}^{2}}{\zeta} \ll 1 $, the transducer is weakly-coupled, and vice versa. For example, consider ME transducers. The tri-layer and bi-layer films in Fig.~\ref{impedance_measurement} have a $\frac{k_{e}^{2}}{\zeta}$ of 2.5 and 8.6, suggesting that they are moderately electromagnetically coupled.

\begin{figure}[!t]
\centering
\includegraphics[width=0.9\linewidth]{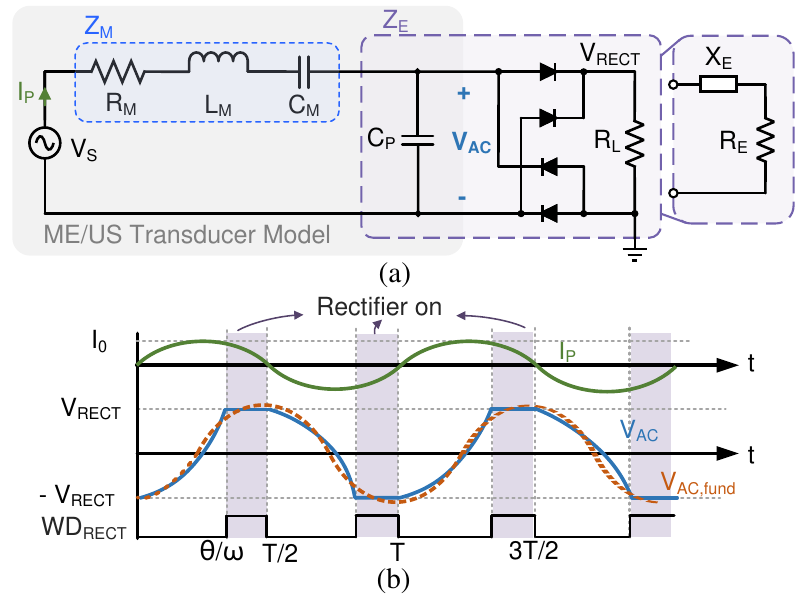}
\vskip -1ex
\caption{(a) ME/US transducer interfacing with a full-bridge rectifier and (b) the associated waveform.}
\label{interface_circuit}
\end{figure}

For ME/US WPT and PEH, an AC-DC converter is connected to the transducer. The most basic and widely used AC-DC is the full-bridge rectifier (FBR). Therefore, we will perform the analysis of the interface circuit based on FBR. Fig.~\ref{interface_circuit} shows the ME/US transducer connecting to a FBR and a load resistor, as well as the associated waveform. The current $I_P$ is the equivalent current generated by the energy source, and the transducer's output voltage $V_{AC}$ is clamped by the rectifier's output voltage $V_{RECT}$. The rectifier watchdog signal (WD\textsubscript{RECT}) indicates the period when the rectifier is on. In \cite{liang_impedance_2012}, an impedance modeling method to transform the nonlinear PEH interface circuit into a linear equivalent impedance is proposed. It simplifies the analysis by making two assumptions: (1) the equivalent current $I_P$ is a sine wave, and (2) only the fundamental tone of $V_{AC}$ affects the output power. The equivalent current can then be expressed as $I_{P}=I_0sin\omega t$. Based on the charging behavior of $C_P$, the voltage $V_{AC}$ can be expressed with a piece-wise function:
\begin{equation}
\label{FBR_VAC}
    V_{AC}(t) = 
    \begin{cases}
        \frac{I_0}{\omega C_{p}}(1-cos\omega t)-V_{RECT}, & 0\leq \omega t<\theta \\
        V_{RECT}, & \theta \leq \omega t < \pi \\
        V_{RECT}- \frac{I_0}{\omega C_{p}}(1+cos\omega t), & \pi\leq \omega t < \theta+\pi \\
        -V_{RECT} & \theta+\pi \leq \omega t < 2\pi
    \end{cases}
\end{equation}
where $V_{RECT}$ can be expressed as a function of $\theta$:
\begin{equation}
	\label{VRECT_eq}
	V_{RECT} = \frac{I_0}{2\omega C_P}(1-cos\theta).
\end{equation}

By taking the Fourier transform of \eqref{FBR_VAC}, we can calculate the fundamental tone:
\begin{equation}
	\label{FBR_VAC_fund}
	V_{AC,fund}(t) = \frac{I_0}{2\pi\omega C_P}[(sin2\theta-2\theta)cos\omega t+2sin^2\theta sin\omega t].
\end{equation}

By dividing $V_{AC,fund}$ with $I_{P}$, the equivalent impedance of the electrical interface is derived:
\begin{equation}
	\label{Ze}
	Z_{E} = \frac{1}{\pi \omega C_P}[sin^2\theta+j(sin\theta cos\theta-\theta)].
\end{equation}

With the equivalent impedance $Z_E$, the current amplitude of $I_P$ can be expressed as

\begin{equation}
	\label{I0}
	I_{0} = \frac{V_{S,amp}}{|Z_M+Z_E|} .
\end{equation}
where $V_{S,amp}$ is the amplitude of the voltage source in the transducer model.

The load $R_L$ can also be expressed as a function of $\theta$ by considering that the charging power during $\theta/\omega <t < T/2$ equals to the average power loss on $R_L$:

\begin{equation}
	\label{RL_eq}
	R_{L} = \frac{\pi}{2\omega C_P} \frac{1-cos\theta}{1+cos\theta}.
\end{equation}

The output power at the rectifier can be calculated as $V_{RECT}^2/R_L$. By substituting $V_{RECT}$, $I_0$ and $R_L$ with \eqref{VRECT_eq}, \eqref{I0} and \eqref{RL_eq}, we can derive the final output power expression:

\begin{equation}
	\label{Pout}
	P_{out} = \frac{V_{RECT}^2}{R_L}= \frac{I_0^2sin^2\theta }{2\pi\omega C_P} = \frac{sin^2\theta}{2\pi\omega C_P}\cdot \frac{V^2_{S,amp}}{|Z_M+Z_E|^2}.
\end{equation}


\subsection{Maximum Power Point of ME/US}

In \eqref{Pout}, the output power of the transducer is expressed as a function of $\theta$, the impedance of the transducer ($Z_M, C_P$), the operating frequency $\omega$, and the input voltage amplitude $V_S$. The parameter $\theta$ is related to the rectification duty cycle ($Duty_{RECT} = 1-\theta /\pi$), which is used to determine the MPP of PEH in \cite{yue_303_2023}. To derive the MPP condition in terms of $Duty_{RECT}$, we take the derivative of $P_{out}$ with respect to $\theta$ and set it to zero. Noting that \eqref{Pout} can be rewritten as 
$P_{out}=f(\theta,Z_M,C_P,\omega)\cdot V^2_{S,amp}$, the MPP condition can be obtained by solving
\begin{equation}
        \label{derivative}
        \begin{split}
            \frac{\text{d}P_{out}}{\text{d}\theta} = \frac{\text{d}f}{\text{d}\theta}\cdot V^2_{S,amp}&=0
    \\
        \text{d}f(\theta,Z_M,C_P,\omega)/{\text{d}\theta}&=0.
        \end{split}
\end{equation}
As shown in \eqref{derivative}, the optimal value of $\theta$ that maximizes $P_{out}$ depends only on the transducer impedance ($Z_M$, $C_P$) and the operating frequency $\omega$, and is independent of the input voltage amplitude $V_S$. This indicates that, for a given transducer, the MPP condition with respect to the rectification duty cycle remains constant regardless of variations in received power. Furthermore, if the model simplifies to the uncoupled case commonly used in piezoelectric energy harvesting, where the current source $I_0$ is constant, \eqref{Pout} reduces to $P_{out}=I^2_0 \cdot sin^2\theta/(2\pi\omega C_P)$. In this case, the optimal $\theta$ is $\pi/2$, corresponding to a 50\% rectification duty cycle, consistent with the result reported in \cite{yue_303_2023}.

\begin{figure}[!t]
\centering
\includegraphics[width=\linewidth]{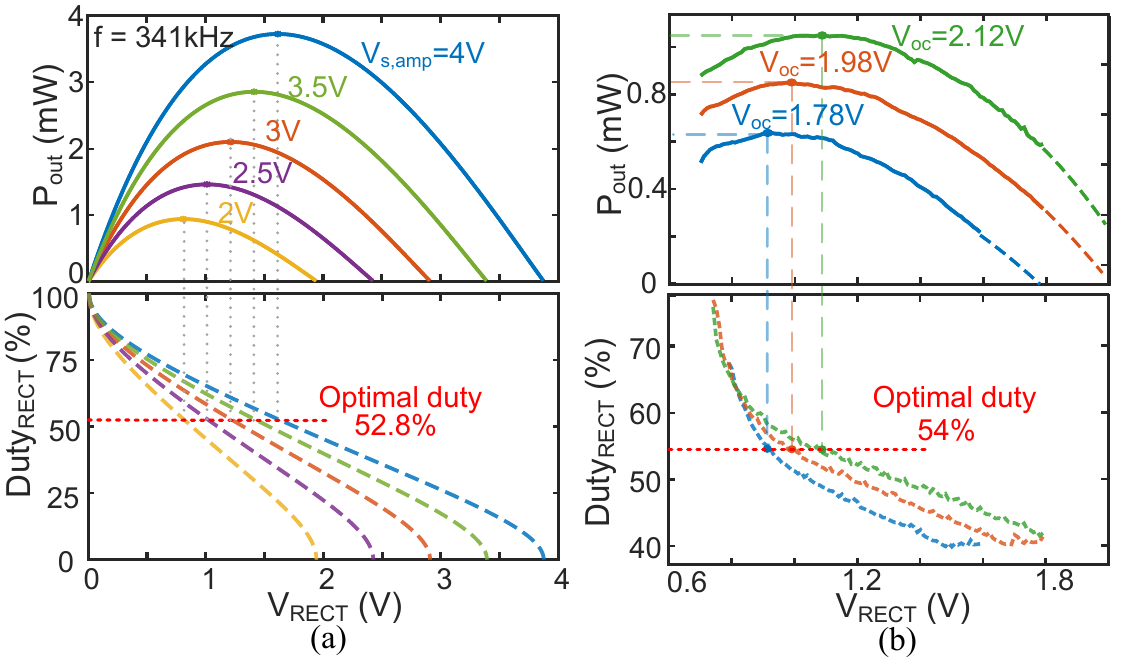}
\vskip -1ex
\caption{Output power and optimal rectification duty cycle results from (a) modeling and (b) measurement.}
\label{Modeled_power_duty}
\end{figure}

\begin{figure}[!t]
\centering
\includegraphics[width=0.85\linewidth]{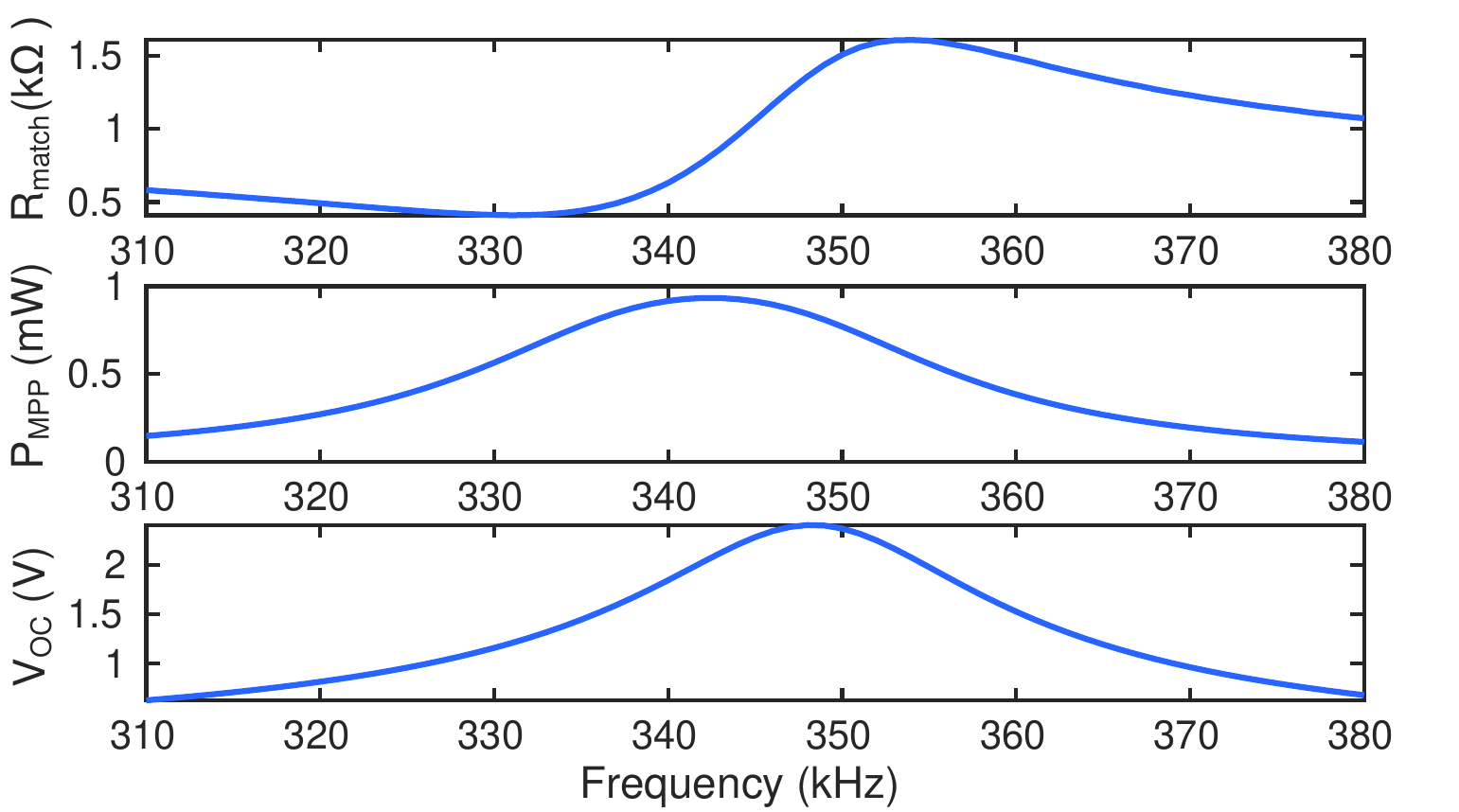}
\vskip -1ex
\caption{Frequency sweep results for the tri-layer ME transducer model, including matched load resistance ($R_{match}$), maximum power point ($P_{MPP}$), and open-circuit voltage ($V_{OC}$).}
\label{P_vs_freq}
\end{figure}

Although deriving a closed-form expression for the optimal $\theta$ from \eqref{derivative} is challenging, a sufficiently accurate solution can be obtained by sweeping $\theta$ from 0 to $\pi$ and identifying the value that maximizes $P_{out}$. Fig.~\ref{Modeled_power_duty}a presents the output power and corresponding optimal rectification duty cycle based on modeling results using the tri-layer ME transducer parameters from Fig.~\ref{impedance_measurement}b. As shown, the optimal duty cycle remains constant across different input amplitudes and is skewed from the 50\% midpoint. Additionally, the $P_{out}$–$V_{RECT}$ relationship deviates from a quadratic curve under the ME model, indicating that the FOCV method (which assumes maximum power at half $V_{OC}$) is not directly applicable, further limiting its suitability for ME transducers.

\begin{figure*}[t!]
\centering
\includegraphics[width=0.8\linewidth]{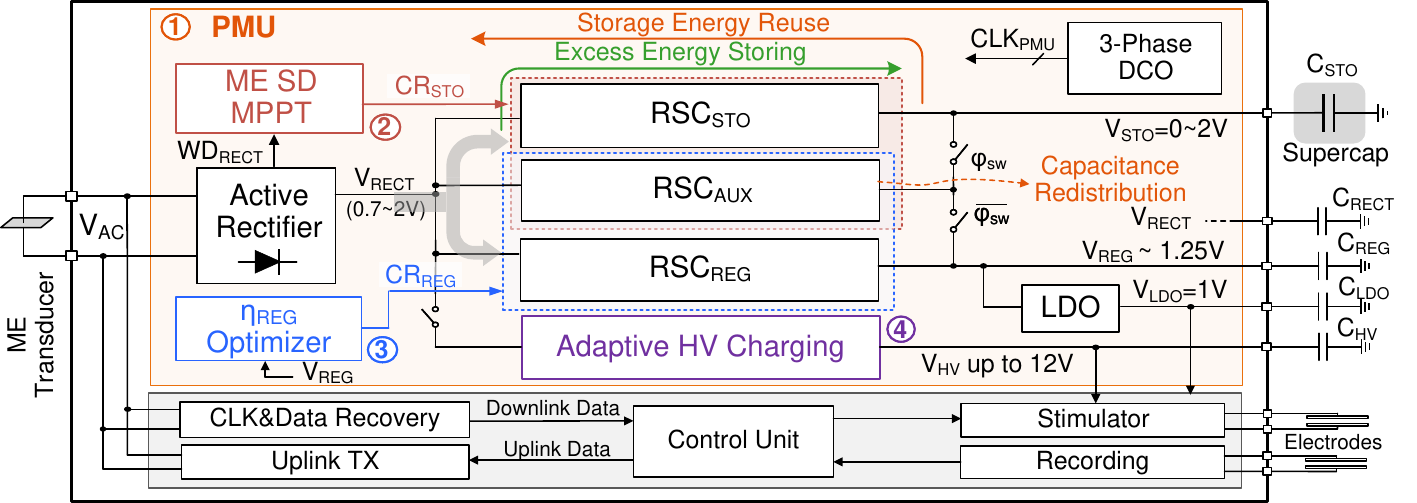}
\caption{System diagram of the presented power management unit embedded in an implant SoC}
\label{System_diagram}
\end{figure*}

The optimal duty cycle was also experimentally measured using the tri-layer ME transducer and the PMU chip, as shown in Fig.~\ref{Modeled_power_duty}b. The Duty\textsubscript{RECT} signal is generated on-chip and monitored via an analog buffer connected to an oscilloscope. 
The measured optimal Duty\textsubscript{RECT} closely aligns with the modeling results, with minor discrepancy. The discrepancy may be introduced by inaccurate transducer model parameters and non-idealities of the on-chip rectifier and the analog buffer used for measurement. While the model does not perfectly match the measured value, it effectively captures the key insight that the optimal duty cycle is largely independent of input power. This confirms that the circuit can be tailored to a specific transducer based on its impedance characteristics. In practical applications with different transducers, a one-time characterization can be performed to determine the optimal duty cycle, which can then be programmed into the PMU accordingly.


Using \eqref{Pout}, we also analyze the impact of carrier frequency on power transfer. Fig.~\ref{P_vs_freq} presents frequency sweep results for the tri-layer ME transducer model, including matched load resistance ($R_{match}$), maximum power point ($P_{MPP}$), and open-circuit voltage ($V_{OC}$). The results reveal that the frequency yielding the highest $P_{MPP}$ differs from that producing the highest $V_{OC}$. This demonstrates that relying solely on $V_{OC}$ to select the frequency may lead to suboptimal performance. Instead, power-based analysis should be performed to guide frequency selection, considering the overall system requirements.

\section{System Design and Circuit Implementation}

\subsection{System Overview}

Fig.~\ref{System_diagram} shows the system architecture of the proposed PMU for ME WPT bio-implants. The input AC power is first converted to DC by an active rectifier, then distributed in parallel to the regulation (REG) and storage (STO) stages, avoiding the cascading loss seen in cascaded power stage topologies \cite{cheng_reconfigurable_2021,li_triple-mode_2017}. Both stages use a reconfigurable switched-capacitor (RSC) converter with 12 conversion ratios. 
To maximize on-chip capacitor usage, a capacitance redistribution topology dynamically assigns the auxiliary (AUX) capacitor bank to either REG or STO based on input power and workload. All three RSC stages are implemented with equal capacitance, based on empirical analysis of both quiescent and active power demands in implant scenarios.

Based on the analysis in Section \ref{sec:analysis}, a continuous skewed-duty-based maximum power point tracking (SD-MPPT) technique is employed. It adjusts the rectification duty cycle (Duty\textsubscript{RECT}) by tuning the STO stage conversion ratio (CR\textsubscript{STO}) to match the ME transducer impedance. 
Unlike the conventional P\&O algorithm, which relies on iterative feedback based on real-time power sensing, the proposed SD-MPPT technique eliminates the need for power monitoring. It uses a fixed optimal duty cycle reference derived from the known behavior of the ME transducer and obtained directly from WD\textsubscript{RECT}, an internally available signal, reducing control complexity and power overhead.

To improve regulation efficiency, a real-time optimizer selects the REG stage conversion ratio (CR\textsubscript{REG}) based on the duty cycle of the regulation enable signal (REG\textsubscript{EN}). 
For high-voltage stimulation, a switched capacitor stimulator (SCS) architecture is implemented \cite{lee_power-efficient_2015}, where the output driver employs an H-bridge constructed by stacking 3V transistors to support the 12V output level \cite{wie_33--11v-supply-range_2024}. To increase the charging efficiency and speed, an adaptive HV charging stage is designed to boost the stimulation capacitor to the target voltage with a tunable conversion ratio.

\begin{figure}[t!]
\centering


\includegraphics[width=0.85\linewidth]{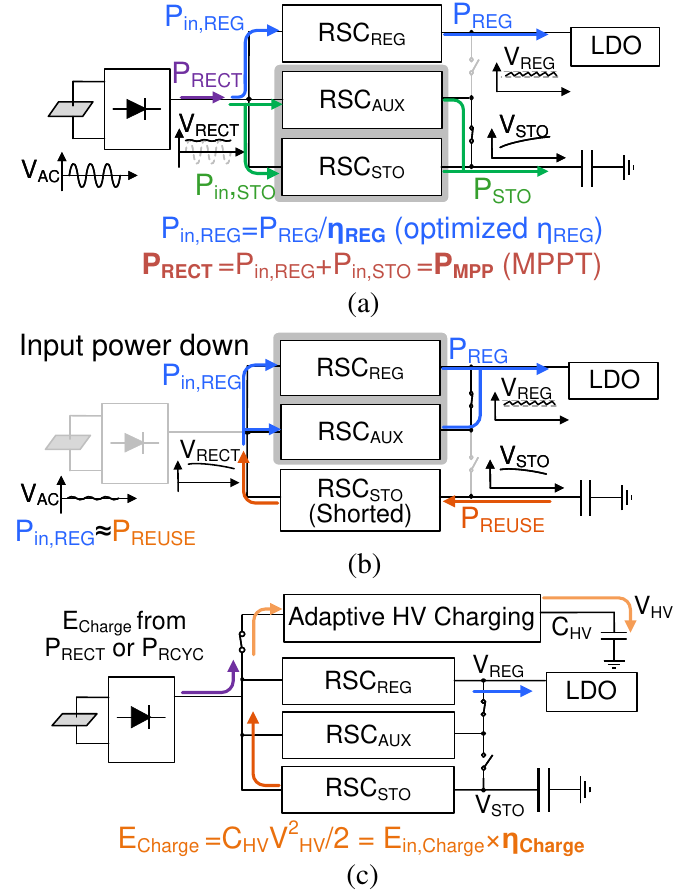}

\caption{Operation modes of the ME WPT PMU. (a) Parallel REG \& STO MPPT mode; (b) Storage reuse mode; (c) High-voltage charging mode. }
\label{operation_modes}
\end{figure}


Fig.~\ref{operation_modes} illustrates the three operation modes of the proposed ME WPT PMU. During normal operation, when input power is present, the rectified power ($P_{RECT}$) is delivered in parallel to the REG and STO stages. The $\eta_{REG}$ optimizer adjusts the REG stage conversion ratio (CR\textsubscript{REG}) to regulate $V_{REG}$ while maximizing efficiency. Once the REG stage draws the minimum input power ($P_{in,REG}$) needed to support the load, the remaining $P_{RECT}$ is directed to the STO stage to charge the supercapacitor. Meanwhile, the SD-MPPT controller tunes CR\textsubscript{STO} to operate the ME transducer at its maximum power point ($P_{MPP}$), maximizing $P_{RECT}$. In this mode, overall system efficiency is optimized and defined as $\eta_{overall} = (P_{REG} + P_{STO}) / P_{MPP}$. 
When input power is low or absent, the system enters storage reuse mode, drawing power from the supercapacitor (maintaining $V_{STO}$) to sustain functionality. In this mode, the power path is reconfigured to draw from stored energy rather than $V_{RECT}$.
During a stimulation task, the high-voltage charging mode is activated to charge the output capacitor. The conversion ratio CR\textsubscript{HV} is adaptively selected based on the output voltage ($V_{HV}$) to improve charging speed and efficiency. HV charging can occur during normal operation using power from the ME transducer or during storage reuse mode using power from the supercapacitor.

\subsection{Skewed-Duty MPPT}

Maximum power point tracking of ME/US transducers can be achieved by adjusting the input impedance of the interface circuit to regulate the rectification duty cycle around a specific target value. This target is determined solely by the transducer’s inherent impedance and remains constant regardless of the input excitation amplitude. In the proposed design, input impedance tuning is realized by adjusting the conversion ratio of the RSC converter. For a given load, increasing the conversion ratio lowers the input impedance of the RSC, thereby enabling impedance matching for optimal power transfer.

\begin{figure}[!t]
\centering
\includegraphics[width=0.9\linewidth]{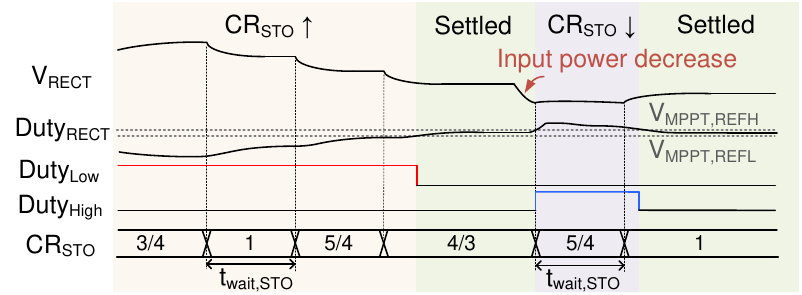}

\caption{Operating waveform of the SD-MPPT for ME WPT.}
\label{SDMPPT_waveform}
\end{figure}

Fig.~\ref{SDMPPT_waveform} shows the operating waveform of the proposed skewed-duty MPPT. During the MPPT phase, CR\textsubscript{STO} is adjusted to keep Duty\textsubscript{RECT} within a reference window. The voltages are generated by an on-chip voltage reference and can be programmed using a digital-to-analog converter (DAC). When Duty\textsubscript{RECT} falls below the lower bound, CR\textsubscript{STO} is increased to lower the input impedance and draw more power from the transducer, which in turn lowers $V_{RECT}$, and vice versa. Since $\eta_{MPPT}$ is insensitive to Duty\textsubscript{RECT} variation near the optimal point, a wider reference window (50\% to 56\%) does not noticeably degrade performance (97\% at the window edges according to measurement) but allows more stable control.
Moreover, to further avoid frequent toggling between two adjacent configurations, the tuning bandwidth of the conversion ratio, set by the tuning interval $t_{wait,STO}$, is placed at a lower frequency, while the other non-dominant poles, such as the pole introduced by the rectifier output capacitor and the transducer, are located far from the dominant pole to minimize unwanted interactions.

\begin{figure}[!t]
\centering
\includegraphics[width=0.9\linewidth]{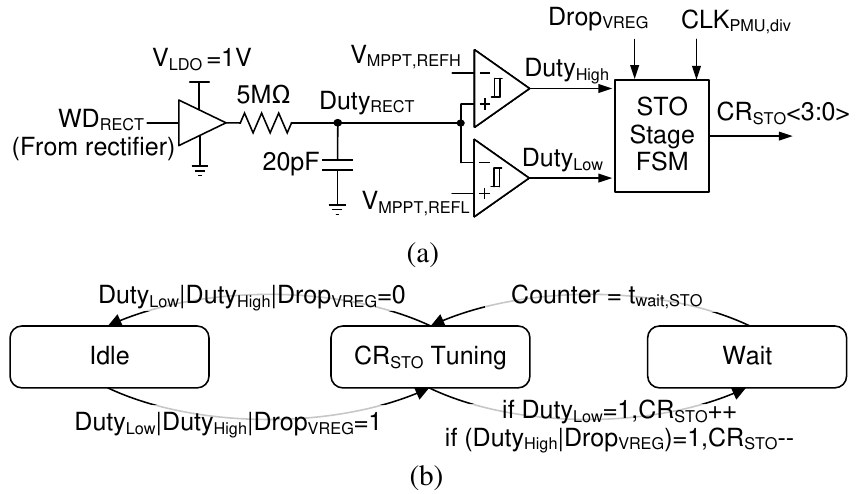}

\caption{(a) Implementation of the SD-MPPT controller and (b) state transfer diagram of the STO stage finite state machine (FSM).}
\label{SDMPPT_circuit}
\end{figure}

Fig.~\ref{SDMPPT_circuit} illustrates the implementation of the MPPT controller. The WD\textsubscript{RECT} signal generated by the active rectifier is digitally buffered with a 1~V supply and filtered with a passive RC into the low-frequency Duty\textsubscript{RECT} signal. Then the Duty\textsubscript{RECT} is compared with two programmable threshold voltages $V_{MPPT,REFH}$ and $V_{MPPT,REFL}$. The comparison results are sent to the FSM to decide CR\textsubscript{STO}.

\subsection{Regulation Efficiency Optimizer}

As the input voltage and output load vary with changes in distance and workload, the regulation stage must adapt its conversion ratio to maintain high efficiency. A common approach involves computing the input-to-output voltage ratio and selecting the corresponding conversion ratio from a look-up table. However, the selected ratio must support the maximum load; otherwise, the regulation stage may fail to maintain the required output voltage. Consequently, the power stage is often over-designed to handle worst-case conditions, resulting in unnecessary area overhead under light-load operation.

To address this, we propose a real-time $\eta_{REG}$ optimizer that dynamically adjusts the conversion ratio of the REG stage to improve efficiency. This method does not rely on prior knowledge of power stage characteristics or load profiles. The design is based on the observation that, in hysteretic regulation, a smaller duty cycle of the regulation enable signal (REG\textsubscript{EN}) at a given load indicates lower efficiency, as more energy is consumed during the active phase of the RSC regulator.

\begin{figure}[!t]
\centering
\includegraphics[width=0.9\linewidth]{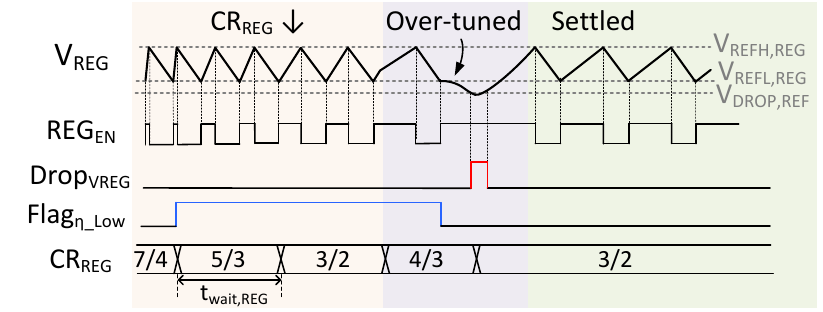}

\caption{Operating waveform of the real-time $\eta_{REG}$ optimizer.}
\label{regulation_optimizer_waveform}
\end{figure}

\begin{figure}[!t]
\centering
\includegraphics[width=0.9\linewidth]{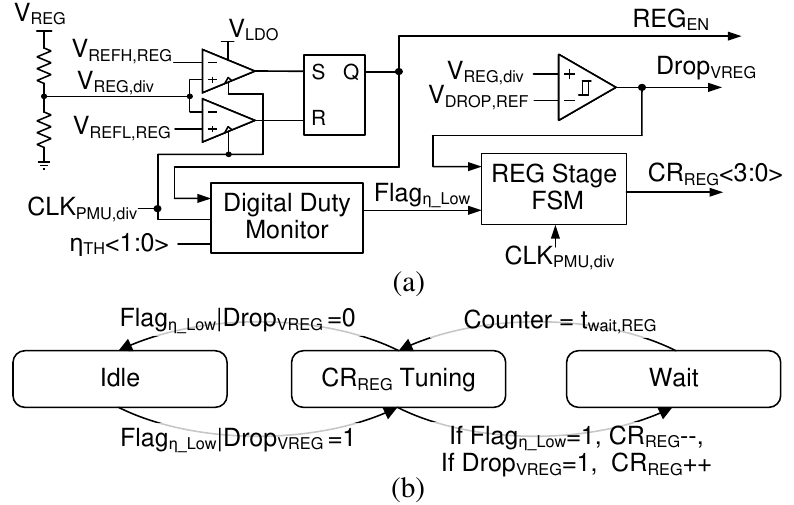}

\caption{(a) Implementation of the $\eta_{REG}$ optimizer and (b) state transfer diagram of the REG stage FSM.}
\label{regulation_optimizer_circuit}
\end{figure}

Fig.~\ref{regulation_optimizer_waveform} illustrates the operating waveform of the $\eta_{REG}$ optimizer. The optimizer increases the duty cycle of REG\textsubscript{EN} by lowering CR\textsubscript{REG} while maintaining regulation. This adjustment is controlled by a digital duty monitor using counters. If CR\textsubscript{REG} is tuned too low and causes $V_{REG}$ to drop, the conversion ratio is incrementally increased until stable regulation is restored. Both the tuning intervals ($t_{wait,REG}$) and the REG\textsubscript{EN} duty cycle threshold are programmable via wireless downlink to ensure stable control of CR.

Fig.~\ref{regulation_optimizer_circuit} shows the implementation of the $\eta_{REG}$ optimizer. Regulation of the $V_{REG}$ rail is achieved by comparing $V_{REG}$, scaled down by half to fit within the $V_{LDO}$ range, against hysteresis voltage references. The comparator output is latched to generate the REG\textsubscript{EN} signal, which enables the REG stage. This signal is also fed to a digital duty monitor, where its duty cycle is compared against a 2-bit programmable threshold ($\eta_{TH}$) using a divided PMU clock to count high and low durations. If the duty cycle falls below the threshold, a Flag\textsubscript{$\eta_low$} signal triggers a finite state machine (FSM) to decrease CR\textsubscript{REG}. Conversely, if $V_{REG}$ drops below the reference $V_{DROP,REF}$, a Drop\textsubscript{VREG} signal prompts the FSM to increase CR\textsubscript{REG}.

\subsection{Adaptive HV Charging}

\begin{figure}[!t]
\centering
\includegraphics[width=\linewidth]{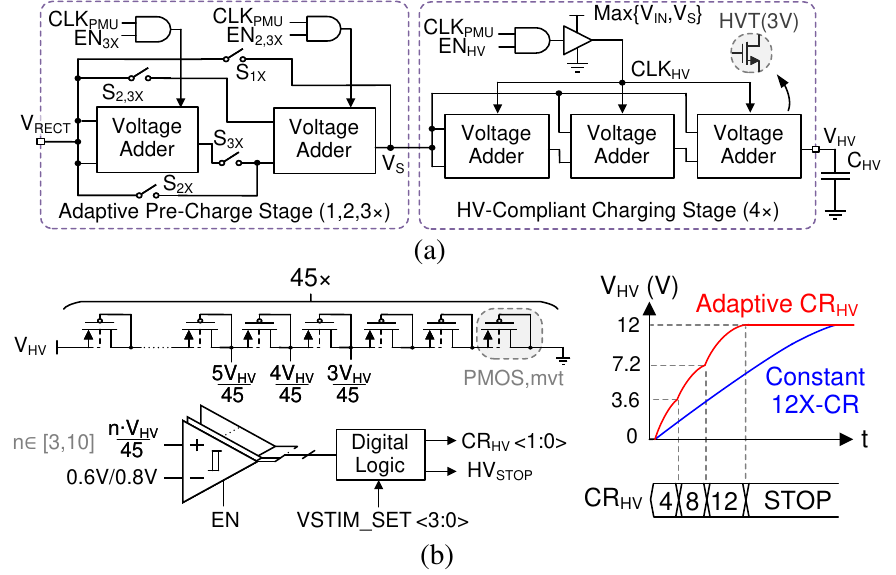}
\vskip -1ex
\caption{(a) HV charging stage topology and (b) control circuit of the adaptive HV charging stage.}
\label{HV_charging_circuit}
\end{figure}


To support high-voltage electrical stimulation under high electrode impedance, a 12~V-compliant adaptive charging stage is integrated into the PMU. As shown in Fig.~\ref{HV_charging_circuit}a, the HV charging stage consists of a pre-charge stage with selectable 1×, 2×, or 3× conversion ratios, followed by a fixed 4× stage implemented using 3~V HVT transistors. The pre-charge conversion ratio is adaptively selected based on the output voltage level, with the target $V_{HV}$ programmable from 2.7~V to 12~V. Adaptive control of CR\textsubscript{HV} and $V_{HV}$ is realized using a PMOS voltage divider and a set of comparators, as shown in Fig.~\ref{HV_charging_circuit}b. To enable fine-grained voltage control while preserving bandwidth, 45 low-threshold PMOS transistors are employed in the divider. Compared to a fixed 12× converter, the proposed adaptive architecture achieves higher efficiency and faster charging speed.

\subsection{Power Stage Design}

\begin{figure}[!t]
\centering
\includegraphics[width=\linewidth]{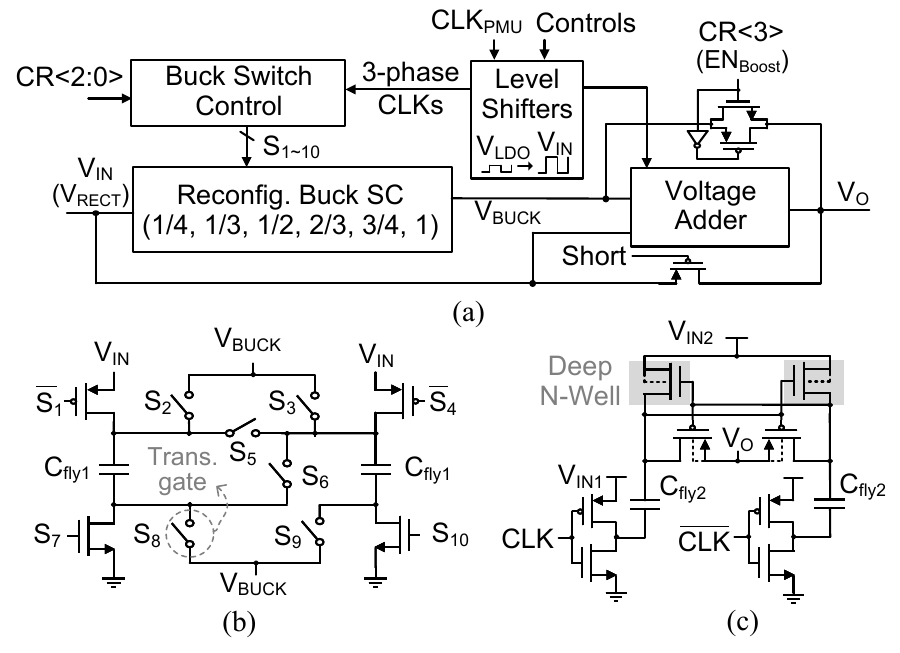}
\vskip -1ex
\caption{Schematic of (a) reconfigurable switched-capacitor (RSC) converter, (b) reconfigurable buck switched-capacitor \cite{jiang_digital_2017}, and (c) cross-coupled voltage adder.}
\label{Power_stage}
\end{figure}

Fig.~\ref{Power_stage} shows the design of one reconfigurable switched-capacitor (RSC) converter. It consists of a 6-ratio reconfigurable buck switched-capacitor converter \cite{jiang_digital_2017} and a cross-coupled voltage adder. 
The detailed switch control implementation of the reconfigurable buck converter can be found in \cite{jiang_digital_2017}.
The flying capacitors are implemented with on-chip metal-insulator-metal (MIM) capacitors with $C_{fly1}=C_{fly2}=83$~pF, which provides sufficient efficiency and load support for our intended use case while keeping the PMU area reasonable. The cross-coupled voltage adder is also used in the adaptive HV charger with different flying capacitor values and transistor sizing.

\begin{figure}[!t]
\centering
\includegraphics[width=0.98\linewidth]{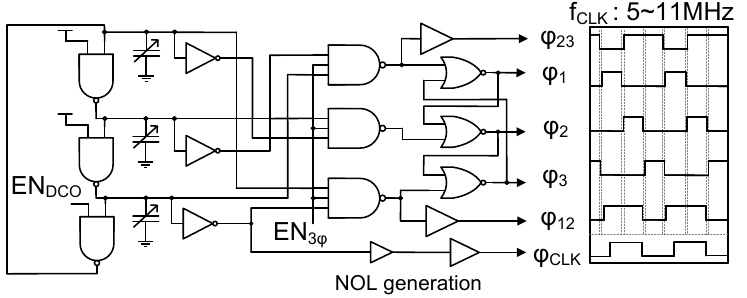}
\vskip -1ex
\caption{Schematic of the 3-phase digitally-controlled oscillator (DCO).}
\label{DCO}
\end{figure}

The operation of the reconfigurable buck SC requires a 2-phase (for CR = 1/3, 1/2, 2/3, and 1) and a 3-phase clock (for CR = 1/4 and 3/4). In this work, a 3-phase digitally-controlled oscillator (DCO) is designed to generate the non-overlapping clocks for the buck SC, as shown in Fig.~\ref{DCO}. The clock frequency is adjustable from 5~MHz to 11~MHz by tuning the capacitor bank in the DCO. The operating frequency selection depends on the throughput power requirement.

\subsection{Startup Circuit}

\begin{figure}[!t]
\centering
\includegraphics[width=0.66\linewidth]{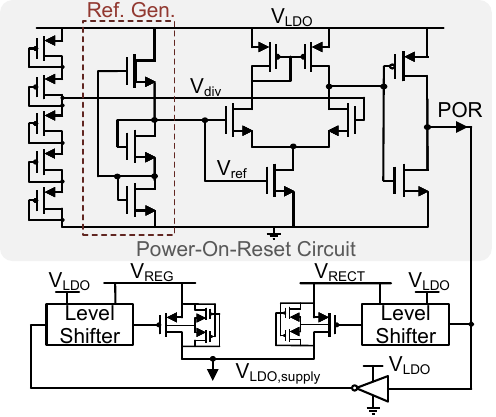}
\vskip -1ex
\caption{Schematic of the startup circuit including the power-on-reset (POR) and the LDO supply switching circuits.}
\label{Startup_circuit}
\end{figure}

To enable reliable cold startup from the harvested ME power without relying on pre-charged storage, a dedicated power-on-reset (POR) startup-assist circuit is implemented, as shown in Fig.~\ref{Startup_circuit}. The POR circuit, comprising a compact subthreshold voltage reference \cite{lee_subthreshold_2017} and a comparator powered by $V_{LDO}$, outputs "0" when $V_{LDO}$ is below approximately 800~mV. This POR signal controls the supply switching circuit, which connects the LDO input to $V_{RECT}$ when POR is low, and switches to $V_{REG}$ once POR goes high. This configuration allows the LDO to be charged up directly from the rectifier during startup, ensuring operation solely from harvested energy without requiring any additional active control.

\section{Measurement Results}

\begin{figure}[!t]
\centering
\includegraphics[width=0.9\linewidth]{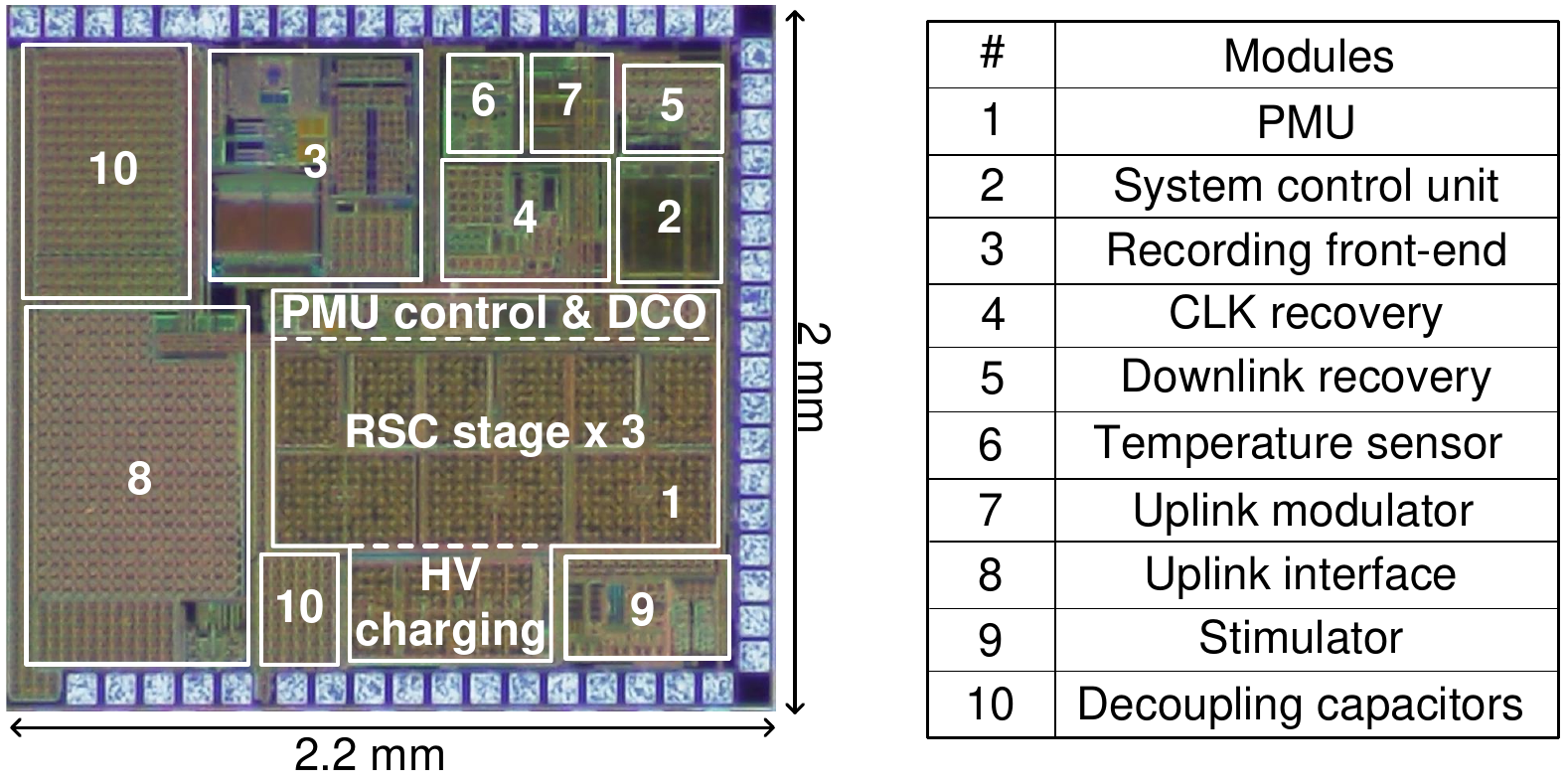}

\caption{Chip micrograph of the implant SoC.}
\label{Die_photo}
\end{figure}

A test chip of the PMU, embedded in a complete System-on-Chip (SoC), was fabricated in 180~nm CMOS technology, as shown in Fig.~\ref{Die_photo}. The SoC is packaged with a $5\times 2\times 0.2~mm^3$ tri-layer ME transducer, a supercapacitor, a 0201 stimulation capacitor, and three 01005 decoupling capacitors. The supercapacitor is placed between two printed circuit board (PCB) layers, and the top surface of the supercapacitor is attached to the upper PCB through silver epoxy for ground connection. The final packaged implantable device occupies a volume of $7\times 3.8\times 2.4~mm^3$ and weighs $106~mg$.

\subsection{System Operation}

\begin{figure}[!t]
\centering
\includegraphics[width=0.95\linewidth]{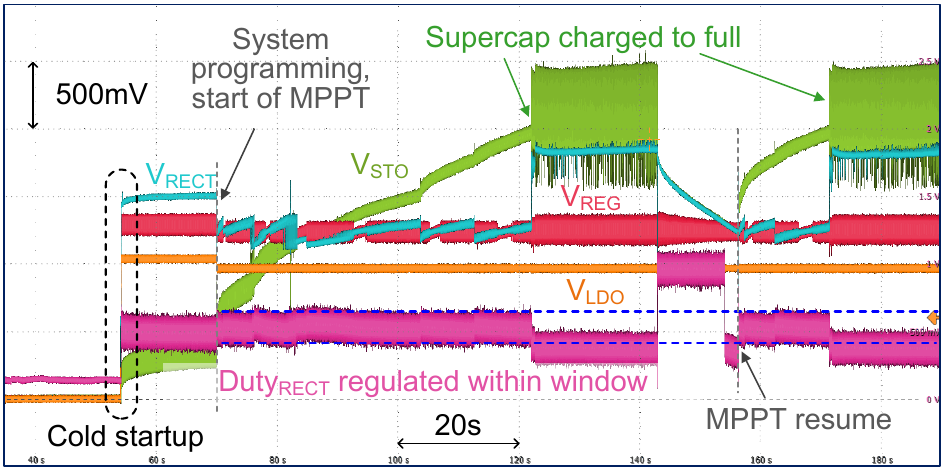}

\caption{Measured system operation waveforms of the presented PMU.}
\label{System_operation_waveform}
\end{figure}

Fig.~\ref{System_operation_waveform} displays the measured system operation waveforms of the PMU. After the system cold-startup and initial programming, the PMU begins charging the supercapacitor while regulating $V_{REG}$ voltage rail. MPPT turns on when the PMU charges $V_{STO}$, during which Duty\textsubscript{RECT} is maintained within the programmed voltage reference window. After $V_{STO}$ is fully charged, the STO stage is paused to prevent overvoltage.

\begin{figure}[!t]
\centering
\includegraphics[width=0.9\linewidth]{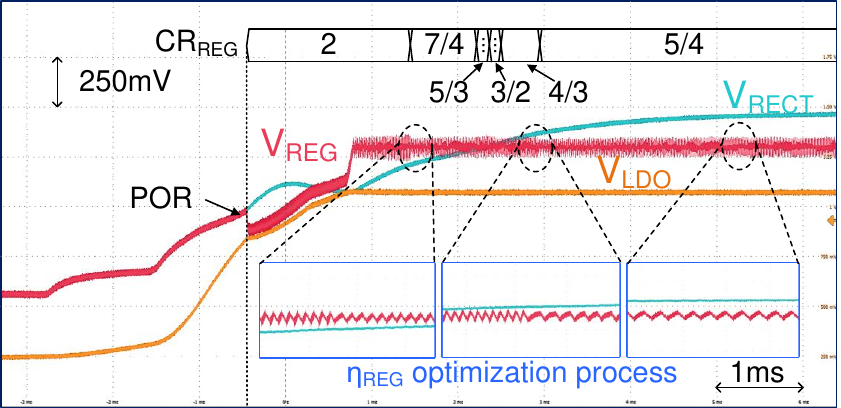}

\caption{Waveforms for cold startup with $\eta_{REG}$ optimization process.}
\label{cold_start}
\end{figure}

A zoomed-in view of the cold startup waveforms in Fig.~\ref{cold_start} illustrates the $\eta_{REG}$ optimization process. As $V_{RECT}$ rises during startup, the $\eta_{REG}$ optimizer compares the duty cycle of the REG\textsubscript{EN} signal with a predefined threshold, and CR\textsubscript{REG} is decreased to meet the REG\textsubscript{EN} duty cycle requirements. Therefore, the efficiency of the regulation stage is enhanced while ensuring $V_{REG}$ remains regulated. It should be noted that the $\eta_{REG}$ optimizer not only adjusts CR\textsubscript{REG} during startup but also operates continuously to monitor and optimize $\eta_{REG}$.

\begin{figure}[!t]
\centering
\includegraphics[width=0.9\linewidth]{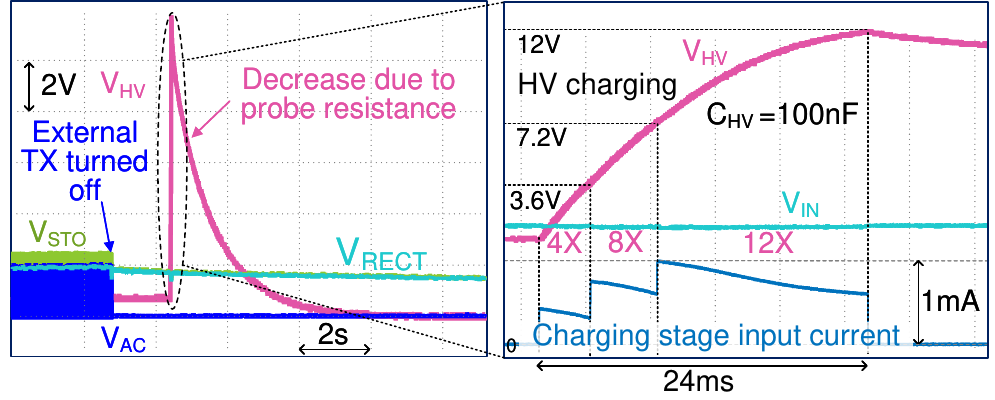}

\caption{Adaptive high-voltage charging after the input power shut-down.}
\label{HV_charging_waveform}
\end{figure}

To demonstrate storage energy reuse and adaptive high-voltage charging, a stimulation task is initiated after shutting down the input power, as shown in Figure~\ref{HV_charging_waveform}. The system detects the power-down event and connects $V_{RECT}$ to the supercapacitor to sustain the system operation and charges the 100~nF stimulation capacitor. $V_{HV}$ is charged to 12~V,  during which the CR\textsubscript{HV} switches from 4× to 8× and 8× to 12× when $V_{HV}$ exceeds 3.6~V and 7.2~V. The charging stage's input current is also monitored during the process, showing that at lower $V_{HV}$, using the smaller CR\textsubscript{HV} can have a smaller input current, therefore has a better charging efficiency.

\begin{figure}[!t]
\centering
\includegraphics[width=0.9\linewidth]{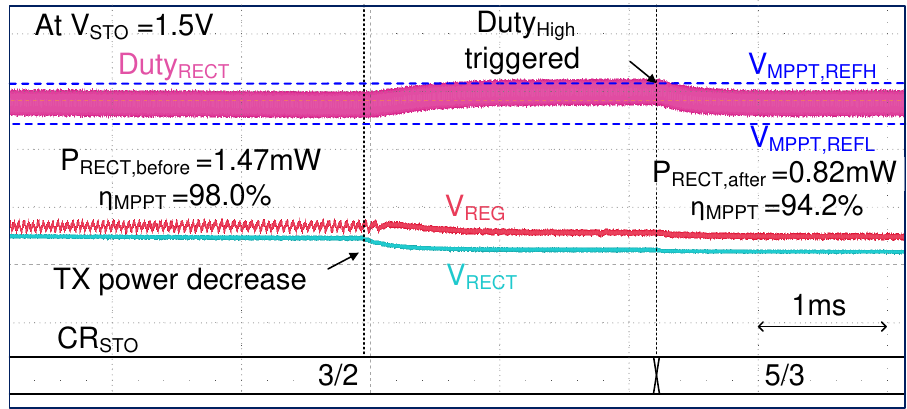}
\caption{MPPT tuning process when the external TX power decreases.}
\label{MPPT_tuning_meas}
\end{figure}

The MPPT tuning process is illustrated in Fig.~\ref{MPPT_tuning_meas}. In this test, the TX power is reduced, which causes the Duty\textsubscript{RECT} to exceed the upper threshold. In response to it, the SD-MPPT module decreases the CR\textsubscript{STO} from 2/3 to 5/3 to take in less power from $V_{RECT}$, making the Duty\textsubscript{RECT} back within the thresholds. The received power at the rectifier's output changes from 1.47~mW to 0.82~mW, corresponding to $\eta_{MPPT}$ of 98.0\% and 94.2\%. 

\subsection{Power Transfer Efficiency}

\begin{figure}[!t]
\centering
\includegraphics[width=\linewidth]{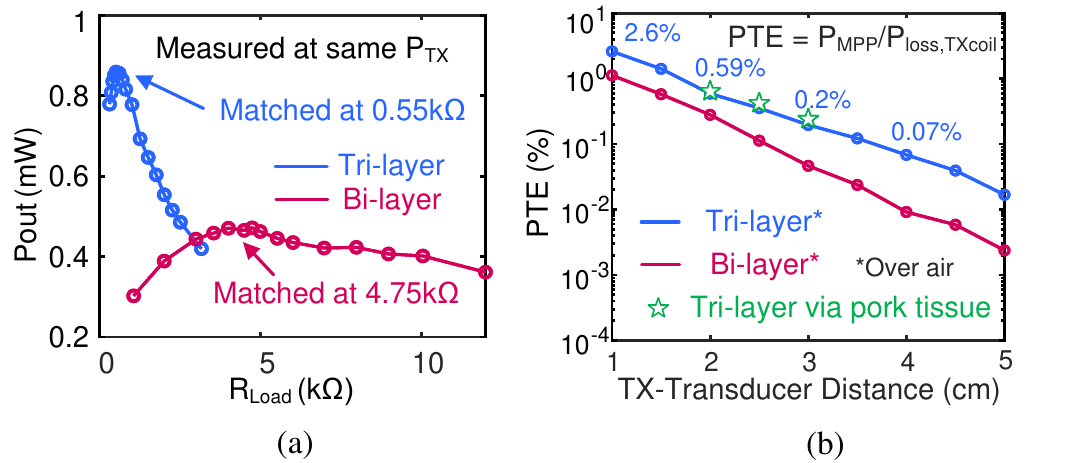}
\vskip -1ex
\caption{(a) Load impedance sweep of the tri-layer and bi-layer ME transducers; and (b) power transfer efficiency across TX-transducer distances.}
\label{PTE}
\end{figure}

To evaluate the performance of the co-optimized ME transducer and PMU, we first measure the output power versus load resistance of tri-layer and bi-layer ME films under the same TX power. As shown in Fig.~\ref{PTE}a, the tri-layer transducer achieves a higher peak power with a matched load of 0.55~k$\Omega$ while the bi-layer transducer matches at 4.75 k$\Omega$. This impedance decrease comes from the thinner PZT layer used in the tri-layer film.  

The power transfer efficiency (PTE), defined as the output power of the transducer at the maximum power point divided by the output power of the TX driver, is measured across varying TX-transducer distances in Fig.~\ref{PTE}b. A PTE of 0.59\% is achieved with the tri-layer ME transducer over the air at 2~cm. The PTE is also tested with pork tissue being the medium. The resulting PTE is slightly higher due to the permeability difference between the pork tissue and the air.

\subsection{MPPT Efficiency}

\begin{figure}[!t]
\centering
\includegraphics[width=\linewidth]{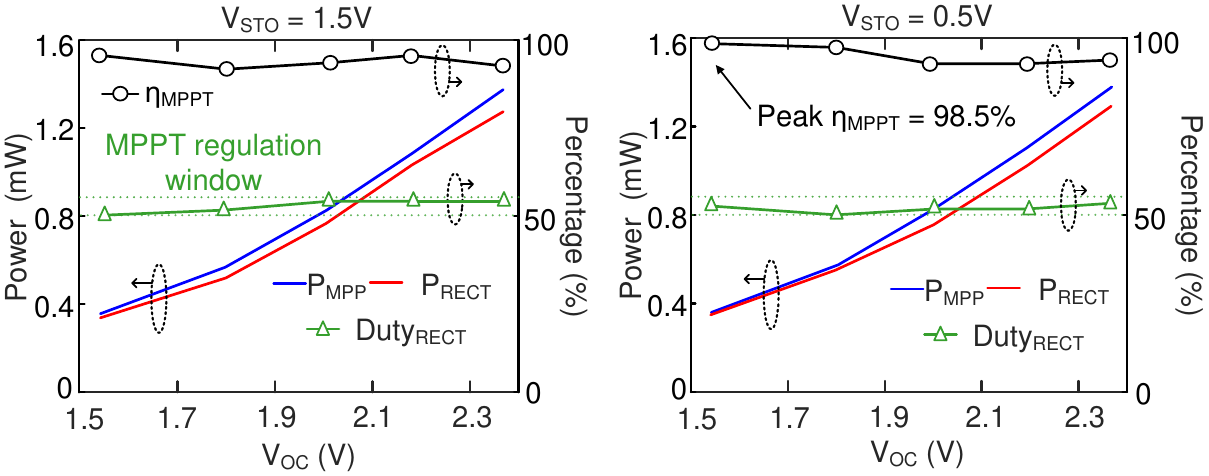}

\caption{Measured MPPT efficiency and the corresponding $P_{RECT}$ and $P_{MPP}$ versus input open-circuit voltage at $V_{STO} = 1.5~V$ and $0.5~V$.}
\label{MPPT_efficiency}
\end{figure}

Fig.~\ref{MPPT_efficiency} presents the measurement results of the maximum power point tracking (MPPT) efficiency ($\eta_{MPPT}$), defined as the ratio of the received power at the rectifier output ($P_{RECT}$) to the maximum available power ($P_{MPP}$). Measurements are conducted under two different $V_{STO}$ conditions (1.5~V and 0.5~V) to emulate varying energy storage states. The maximum available power under each input condition is determined by sweeping the output load and identifying the peak power point. The PMU demonstrates consistently high MPPT efficiency across a range of input power levels and output voltages, achieving a peak $\eta_{MPPT}$ of 98.5\%. $\eta_{MPPT}$ deviations from the peak are attributed to the use of discrete conversion ratios, which limit impedance matching resolution, and a relaxed Duty\textsubscript{RECT} window that trades off tracking accuracy for control stability.

\subsection{System Overall Efficiency}

\begin{figure}[!t]
\centering
\includegraphics[width=\linewidth]{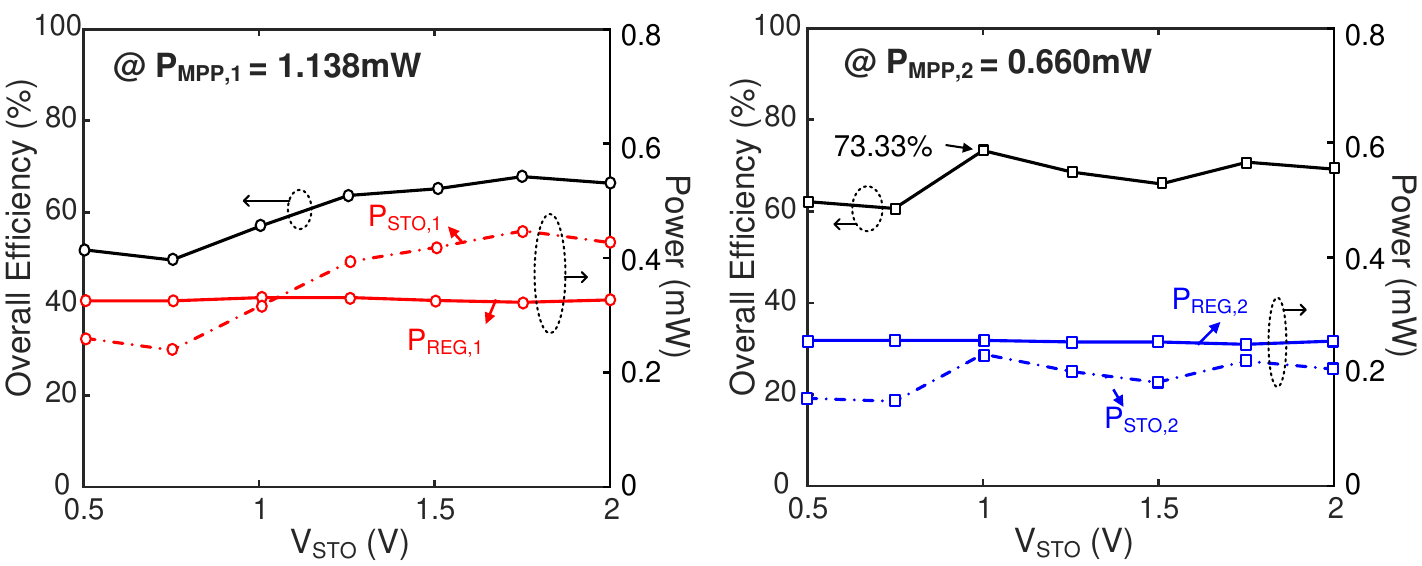}

\caption{Measured system overall efficiency at two input power levels ($P_{MPP}$=1.138~mW and 0.660~mW).}
\label{System_overall_efficiency}
\end{figure}

In this work, the system overall efficiency ($\eta_{overall}$) is defined as the summation of the output powers from the REG and STO stages divided by the maximum available power, expressed as $(P_{REG} + P_{STO}) / P_{MPP}$. This metric reflects the amount of input power that is effectively utilized by the PMU, accounting for both the power delivered to the load and the energy stored. Fig.~\ref{System_overall_efficiency} shows the measured $\eta_{overall}$ as a function of the storage stage output voltage, under two different input power levels ($P_{MPP}$ = 1.138~mW and 0.660~mW). The PMU maintains over 50\% efficiency across all tested conditions, achieving a peak $\eta_{overall}$ of 73.33\%.

\subsection{Power Conversion Efficiency}

\begin{figure}[!t]
\centering
\includegraphics[width=\linewidth]{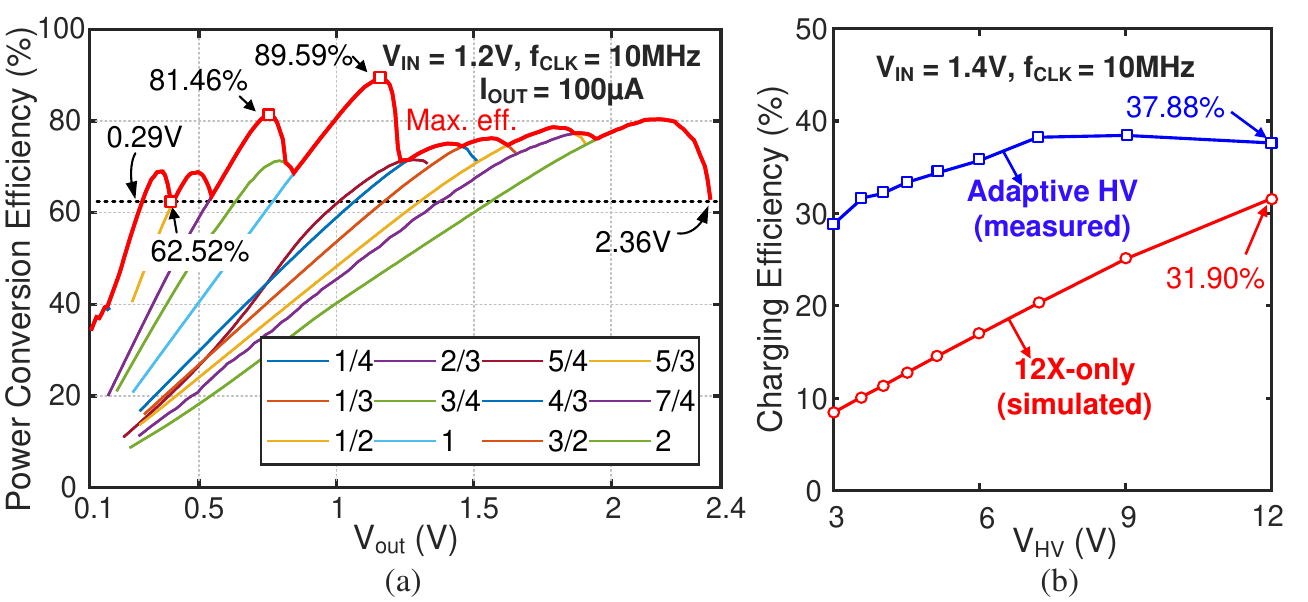}
\vskip -1ex
\caption{(a) Measured power conversion efficiency of a single RSC power stage and (b) HV charging efficiency.}
\label{PCE_charging}
\end{figure}


To characterize the RSC power stage design, the power conversion efficiency of a single RSC stage is measured, as shown in Fig.~\ref{PCE_charging}a. The measurement is performed with an input voltage of 1.2~V and a switching clock frequency of 10~MHz. The input and output power values are obtained using sourcemeters. The RSC stage achieves over 62.52\% efficiency over a wide output range from 0.29~V to 2.36~V. The peak efficiencies of 81.46\% and 89.59\% are achieved with the conversion ratios of 2/3× and 1×. 
The main power losses of the RSC come from charge redistribution/switch conduction, parasitic capacitance of the on-chip flying capacitors, and switching control overhead from the decoder and level shifters.

\subsection{HV Charging Efficiency}


Fig.~\ref{PCE_charging}b illustrates the adaptive HV charging efficiency, defined as the ratio of energy stored in the output capacitor to the total input energy consumed by the charging stage. During the measurement, a sourcemeter supplies the input voltage and simultaneously monitors the input current, while the output voltage is captured using an oscilloscope. For comparison, a fixed 12× charging stage is simulated, with all parasitic effects included. As shown in the results, the adaptive HV charging stage consistently achieves higher efficiency across a wide range of stimulation voltages (3~V to 12~V), reaching up to 37.88\% efficiency when charging to 12~V, outperforming the fixed 12× design.

\subsection{Comparison}

Table~\ref{table1} compares our ME WPT PMU with state-of-the-art mm-scale ME/US-powered bio-implants. Thanks to the SD-MPPT and supercapacitor storage, this work receives load-independent maximum available power from the tri-layer transducer. With the regulation efficiency optimizer, the PMU adaptively maintains high overall efficiency under varying input and load conditions.

\begin{table*}[]
\caption{\textbf{Comparison with mm-Scale ME/US Bio-Implants}}
\label{table1}
\setlength{\tabcolsep}{4pt}
\begin{tabular}{|c|c|c|c|c|c|}

\hline
\multicolumn{1}{|l|}{} & \textbf{This Work}                                                                               & \begin{tabular}[c]{@{}c@{}}Z. Yu, \\ ISSCC'24\cite{yu_millimetric_2024} \end{tabular}               & \begin{tabular}[c]{@{}c@{}}J. Chen, \\ Nat. BME'20\cite{chen_wireless_2022}\end{tabular} & \begin{tabular}[c]{@{}c@{}}D. Piech, \\ Nat. BME'20\cite{piech_wireless_2020}\end{tabular} & \begin{tabular}[c]{@{}c@{}}S. Sonmezoglu, \\ ISSCC'20\cite{sonmezoglu_45mm3_2020}\end{tabular} \\ \hline
Power Source & \textbf{ME (0.35MHz)} & ME (0.33MHz) & ME (0.34MHz) & Ultrasonic (1.85MHz) & Ultrasonic   (2MHz) \\ \hline
Transducer Size (mm) & \textbf{5x2x0.2 (Tri-layer)} & 5x2x0.2 (Bi-layer) & 5x1.75x0.3 (Bi-layer) & 1x0.8x0.8 & 0.75x0.75x0.75 \\ \hline
Power Management & \textbf{\begin{tabular}[c]{@{}c@{}}Rectifier  + Parallel-input \\ reconfig. SC + LDO\end{tabular}} & \begin{tabular}[c]{@{}c@{}}Rectifier + \\ Adaptive SC + LDO\end{tabular} & Rectifier + LDO & Rectifier + LDO & Rectifier + LDOs \\ \hline

\begin{tabular}[c]{@{}c@{}}Load-Independent Max. Power\end{tabular} & \textbf{Y (MPPT+P\textsubscript{excess} storage)}& N& N& N& N \\ \hline
Power Transfer Efficiency & \textbf{0.59\% (20mm)} & 0.37\% (20mm)                                                            & 0.12\% (20mm) & 0.06\% (18mm) & N/R 
\\ \hline
\begin{tabular}[c]{@{}c@{}}Peak PMU Overall Efficiency\end{tabular} & \textbf{73.33\%}& N/R & N/R & N/R & N/R \\ \hline
\end{tabular}
\end{table*}

\begin{table*}[]
\caption{\textbf{Comparison with State-of-the-art PMUs for IoT and IMD}}
\label{table2}
\setlength{\tabcolsep}{2.5pt}
\begin{tabular}{|c|c|c|c|c|c|c|c|c|}
\hline
 &
  \textbf{This Work} &
  \begin{tabular}[c]{@{}c@{}}H.-C Cheng,\\ JSSC'21\cite{cheng_reconfigurable_2021}\end{tabular} &
  \begin{tabular}[c]{@{}c@{}}H. Kim,\\ JSSC'21\cite{kim_dual-mode_2021}\end{tabular} &
  \begin{tabular}[c]{@{}c@{}}A. Talkhooncheh,\\ JSSC'21\cite{talkhooncheh_biofuel-cell-based_2021}\end{tabular} &
  \begin{tabular}[c]{@{}c@{}}J. Li,\\ JSSC'17\cite{li_triple-mode_2017}\end{tabular} &
  \begin{tabular}[c]{@{}c@{}}W. Jung,\\ ISSCC'16\cite{jung_85_2016}\end{tabular} &
  \begin{tabular}[c]{@{}c@{}}X.Yue,\\ ISSCC'23\cite{yue_303_2023}\end{tabular} &
  \begin{tabular}[c]{@{}c@{}}H.-S. Lee,\\ ISSCC'24\cite{lee_273_2024}\end{tabular} \\ \hline
  
Technology &
  \textbf{180   nm} &
  180   nm &
  180   nm &
  65   nm &
  65   nm &
  180   nm &
  180   nm &
  250   nm \\ \hline
  
Applications &
  \begin{tabular}[c]{@{}c@{}} \textbf{ME WPT}, \\ \textbf{implant} \end{tabular}&
  \begin{tabular}[c]{@{}c@{}} PV EH, \\IoT \end{tabular}&
  \begin{tabular}[c]{@{}c@{}} TEG EH, \\IoT \end{tabular}&
  \begin{tabular}[c]{@{}c@{}} BFC EH, \\sensor \end{tabular} &
  \begin{tabular}[c]{@{}c@{}} PV EH, \\IoT \end{tabular}&
  \begin{tabular}[c]{@{}c@{}} Battery, \\IoT \end{tabular}&
  \begin{tabular}[c]{@{}c@{}} PEH, \\IoT \end{tabular}&
  \begin{tabular}[c]{@{}c@{}} Inductive, \\implant \end{tabular} \\ \hline
\multirow{2}{*}{Converter Topology} &
  \textbf{Buck-Boost} &
  Boost &
  Boost &
  Boost,   Buck &
  Boost,   Buck &
  Buck-Boost &
  Buck-Boost &
  Rectifier \\ \cline{2-9} 
 &
  \textbf{\begin{tabular}[c]{@{}c@{}}Reconfig. \\ SC \\ (12-ratio)\end{tabular}} &
  \begin{tabular}[c]{@{}c@{}}Reconfig. \\ SC \\ (3-ratio)\end{tabular} &
  \begin{tabular}[c]{@{}c@{}}CSCR + \\ Reconfig. SC\\ (2/3X,2X)\end{tabular} &
  \begin{tabular}[c]{@{}c@{}}Reconfig.  \\ SC  \\ (12-ratio)\end{tabular} &
  \begin{tabular}[c]{@{}c@{}}2,3X   Boost+\\ 6$\sim$9X Boost+\\ 1/4X Buck\end{tabular} &
  \begin{tabular}[c]{@{}c@{}}7b Binary+\\ 2X Boost+\\ 1/3X Buck\end{tabular} &
  Inductive &
  SIMO  R\textsuperscript{3} \\ \hline
Cascading-Loss Free\textsuperscript{*} &
  \textbf{Y} &
  N &
  N &
  N &
  N &
  N/A &
  N/A &
  Y \\ \hline
\begin{tabular}[c]{@{}c@{}}Fully-Integrated \\Power Stage\end{tabular} &
  \textbf{Y} &
  Y &
  Y &
  Y &
  Y &
  Y &
  Inductor &
  LC   tank \\ \hline
Capacitor Redistribution &
  \textbf{Y} &
  Y &
  Y &
  N &
  N &
  N &
  N/A &
  N/A \\ \hline
On-chip C\textsubscript{fly} &
  \textbf{1.18  nF} &
  2.2  nF &
  10.2  nF &
  N/R &
  0.6   nF &
  3   nF &
  N/A &
  N/A \\ \hline
MPPT Scheme &
  \textbf{Skewed Duty} &
  FOCV &
  SC  PFM &
  2D Hill-Climbing &
  Off-chip &
  N/A &
  Duty  Cycle &
  N/A \\ \hline
Peak $\eta_{MPPT}$ &
  \textbf{98.50\%} &
  N/R &
  99.80\% &
  N/R &
  N/R &
  N/A &
  98\% &
  N/A \\ \hline
V\textsubscript{in} (V) &
  \textbf{0.7-2} &
  0.45-0.9 &
  0.1-0.5 &
  0.25-1 &
  0.25-0.45 &
  0.9$\sim$4 &
  0$\sim$5 &
  4.8 \\ \hline
V\textsubscript{out} (V) &
  \textbf{0-2} &
  1.5 &
  0.75/1.2-1.45 &
  0.9-1.5 &
  3/0.45 &
  0.6/1.2/3.3 &
  0$\sim$5 &
  1-3.5/4.5 \\ \hline
Throughput Power (mW) &
  \textbf{\textless{}3.87\textsuperscript{***}} &
  \textless{}3 &
  \textless{}20.8 &
  \textless{}0.1 &
  \textless{}0.106 &
  \textless{}0.5 &
  N/R &
  135.53 \\ \hline
\begin{tabular}[c]{@{}c@{}}Power/PMU Area\textsuperscript{**} \\ (mW/mm\textsuperscript{2})\end{tabular} &
  \textbf{3.42} &
  1.77 &
  5.34 &
  0.07 &
  0.22 &
  0.212 &
  N/A &
  N/A \\ \hline
Peak PCE &
  \textbf{\begin{tabular}[c]{@{}c@{}}81.46\%(2/3X)\\ 89.59\%(1X)\end{tabular}} &
  \begin{tabular}[c]{@{}c@{}}69.5\%(2X)\\ 65.7\%(1.5X)\end{tabular} &
  85.40\% &
  86\% &
  \begin{tabular}[c]{@{}c@{}}63.8\%\\ (Step-up SC)\end{tabular} &
  81\% &
  N/R &
  90.82\% \\ \hline
Area (mm\textsuperscript{2}) &
  \textbf{1.13} &
  1.69 &
  3.89 &
  1.4 &
  0.48 &
  2.36 &
  0.47 &
  1.56 \\ \hline
\end{tabular}

\begin{tablenotes}
\footnotesize
\item[]
{N/A: Not available; N/R: Not reported;  $^*$ Only through one converter from input to regulated rails or from input to storage;} \\
$^{**}$ For fully-integrated designs, defined as throughput power/active area; $^{***}$ Max power with $>$70\% efficiency.

\end{tablenotes}

\end{table*}
We also compare our PMU with state-of-the-art PMUs for IMD and Internet-of-Things (IoT) applications in Table~\ref{table2}. Because of the parallel architecture of the regulation and storage stages, our PMU avoids the cascading power loss. In addition, the proposed SD-MPPT for ME WPT achieves comparable peak MPPT efficiency to other works with MPPT.

\section{Conclusion}


In conclusion, we present an adaptive and efficient power management unit (PMU) specifically designed for millimetric bio-implants wirelessly powered by magnetoelectrics (ME). To mitigate temporary power interruptions caused by body motion and harvest excess energy, a miniature supercapacitor is integrated as an energy buffer. A skewed-duty-cycle MPPT technique is introduced to continuously extract maximum power from the transducer under varying input and workload conditions. Additionally, a regulation efficiency optimizer, based on the regulation duty cycle, dynamically adjusts the conversion ratio to maximize efficiency. An adaptive high-voltage charging stage is also implemented, capable of delivering up to 12~V with improved speed and efficiency. Integrated within a complete SoC, the proposed PMU achieves a peak MPPT efficiency of 98.5\% and an overall system efficiency of 73.33\%, demonstrating its suitability for millimeter-scale bio-implant applications.


\ifCLASSOPTIONcaptionsoff
  \newpage
\fi

\bibliography{bib/bibliography_no_note, bib/bstcontrol}
\bibliographystyle{IEEEtran}

\vskip -1\baselineskip
\begin{IEEEbiography}[{\includegraphics[width=1in,height=1.25in,clip,keepaspectratio]{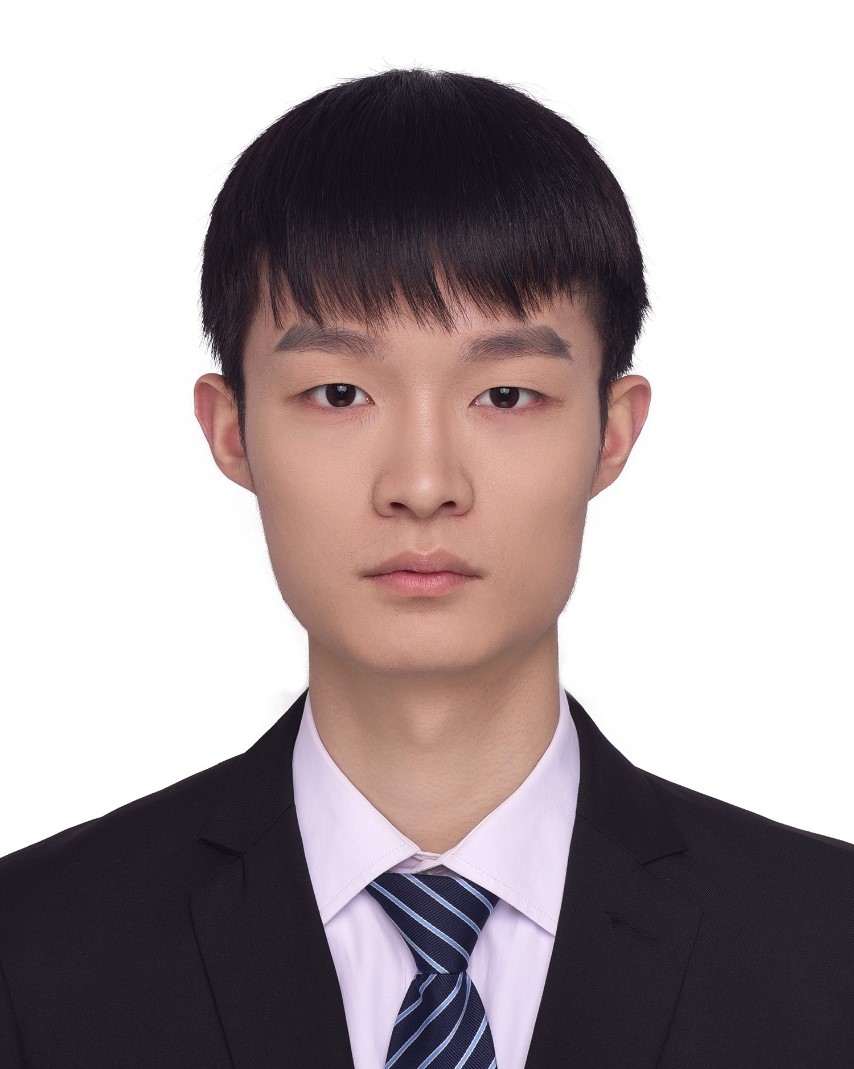}}]{Yiwei Zou} (Graduate Student Member, IEEE) received the B.E. degree in Integrated Circuits and Systems from Huazhong University of Science and Technology, Wuhan, China, in 2022. He is currently working toward his Ph.D. degree in Electrical and Computer Engineering at Rice University, Houston, TX. His research interests include power management and analog/mixed-signal integrated circuits for bioelectronics.
\end{IEEEbiography}

\begin{IEEEbiography}[{\includegraphics[width=1in,height=1.25in,clip,keepaspectratio]{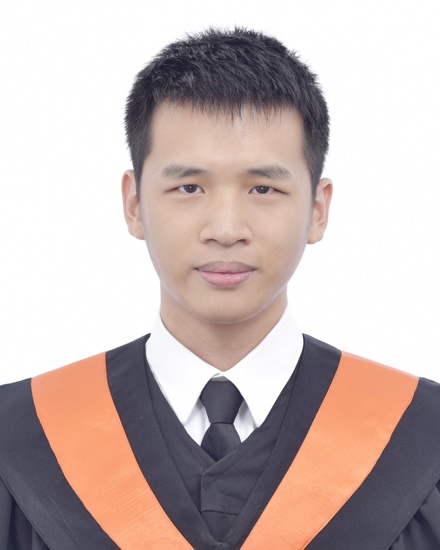}}]{Huan-Cheng Liao} (Graduate Student Member, IEEE) received the B.S. degree in Engineering Science from National Taiwan University, Taiwan, in 2020. He is currently working toward his Ph.D. degree in Electrical and Computer Engineering at Rice University, Houston, TX, USA. His research interests include analog and mixed-signal integrated circuits design.
\end{IEEEbiography}

\vskip -1\baselineskip
\begin{IEEEbiography}[{\includegraphics[width=1in,height=1.25in,clip,keepaspectratio]{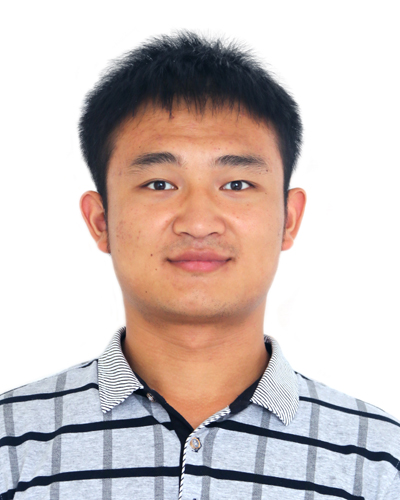}}]{Wei Wang} (Graduate Student Member, IEEE) received a B.S. degree in electronic information science and technology from the Harbin Institute of Technology, Harbin, China, in 2016, and an M.S. degree in integrated circuit engineering from Tsinghua University, Beijing, China, in 2019. He is currently pursuing a Ph.D. degree in electrical
and computer engineering at Rice University, Houston, TX, USA.
His research interests include mixed-signal circuits and systems design.
\end{IEEEbiography}

\vskip -1\baselineskip
\begin{IEEEbiography}
[{\includegraphics[width=1in,height=1.25in,clip,keepaspectratio]{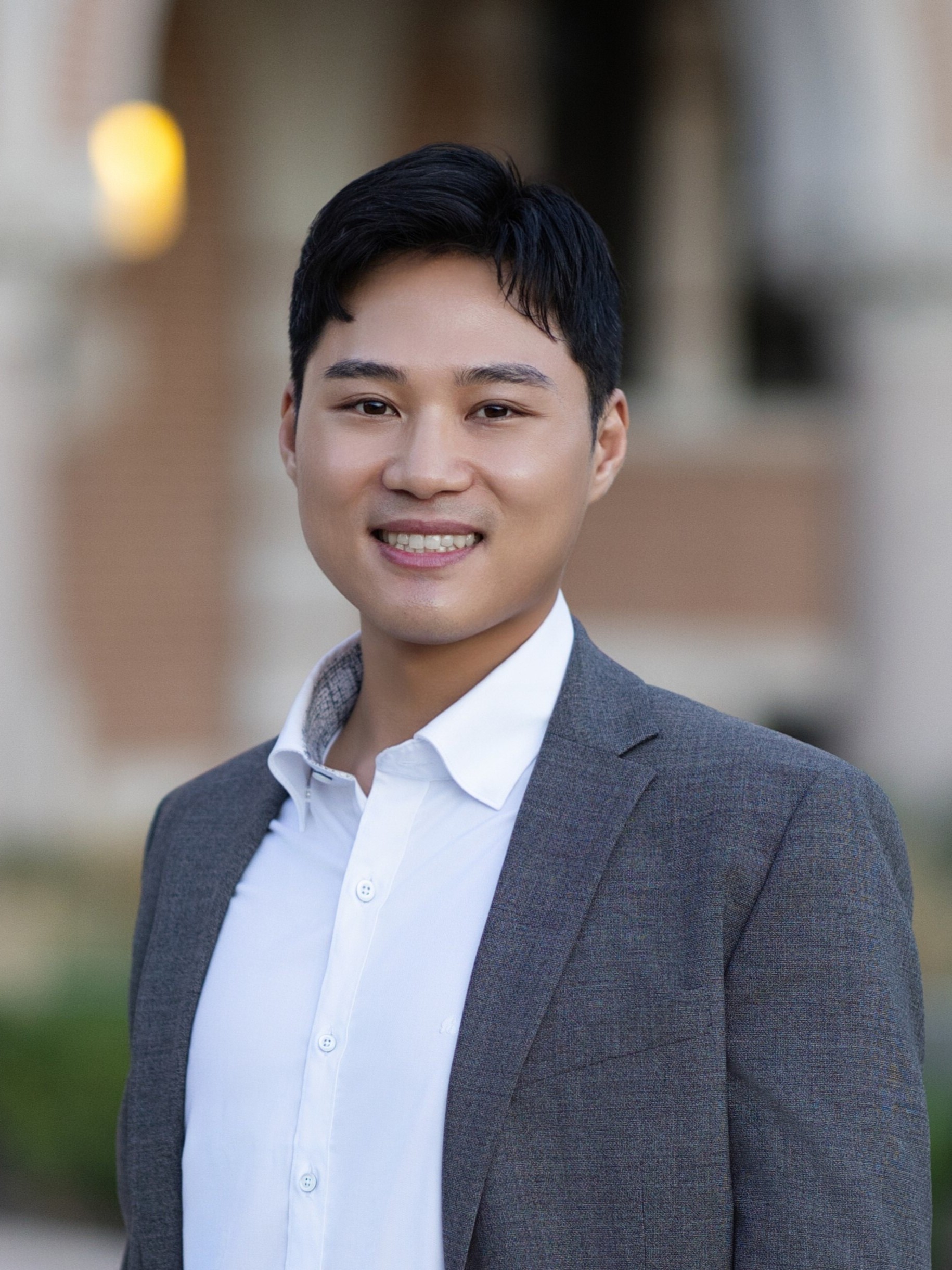}}]{Wonjune Kim} (Graduate Student Member, IEEE)  is a Ph.D. candidate in Electrical and Computer Engineering at Rice University, Houston, TX. They received the M.S. in Bio-convergence Engineering in 2020 and the B.S. in Biomedical Engineering in 2018, both from Korea University, Seoul. Their research focuses on implantable bioelectronics, including magnetoelectric wireless power and data links for mm-scale bio-implants, with interests spanning neuroengineering, embedded systems, and micro/nano-fabrication.
\end{IEEEbiography}

\vskip -1\baselineskip
\begin{IEEEbiography}
[{\includegraphics[width=1in,height=1.25in,clip,keepaspectratio]{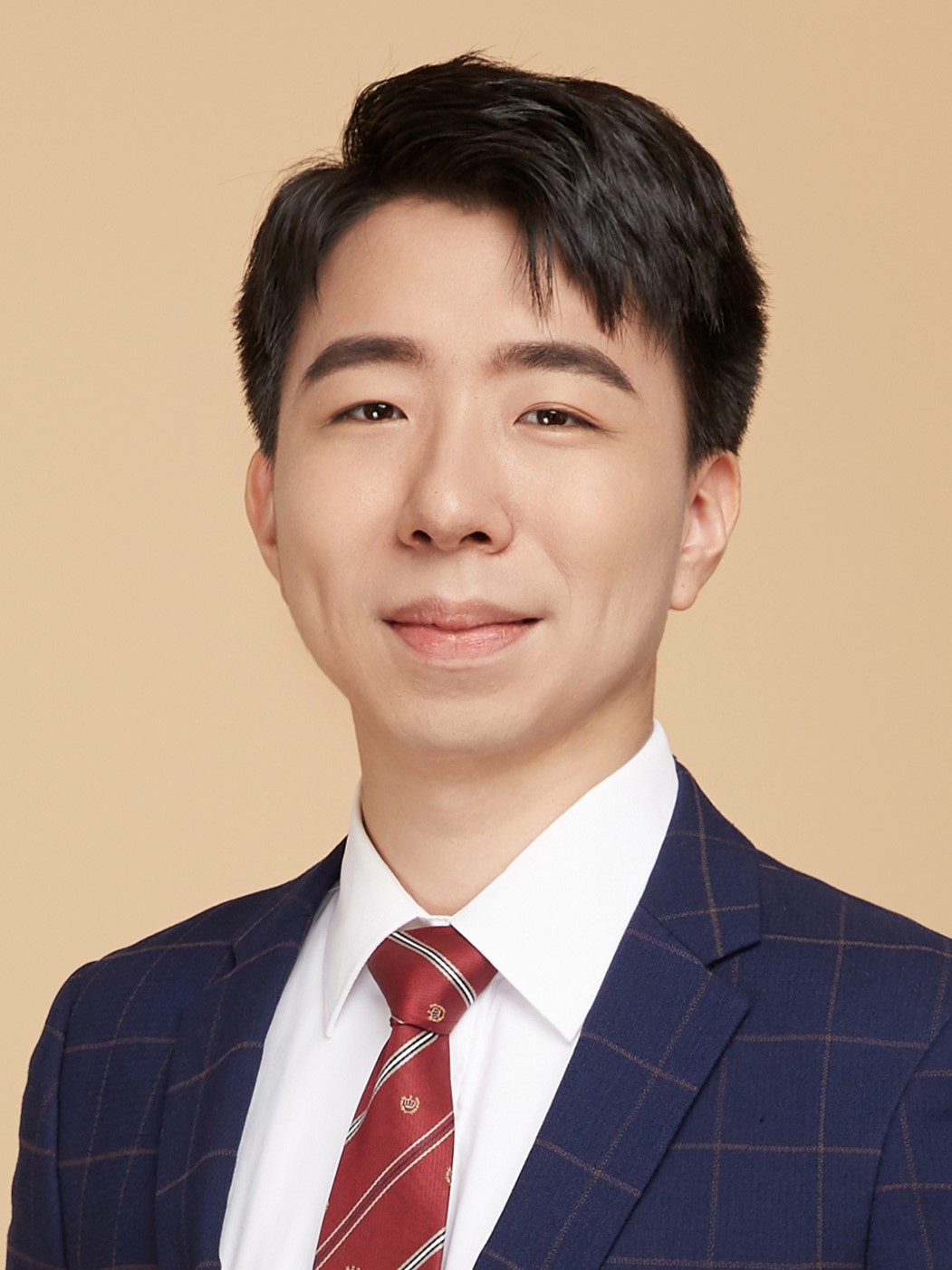}}]{Yumin Su} (Graduate Student Member, IEEE) received the B.S. degree in electrical and computer engineering from Rice University, Houston, TX, USA, in 2023. He is currently pursuing a Ph.D. degree in electrical and computer engineering at Rice University, Houston, TX, USA. 
His research interests include hardware security and design automation.
\end{IEEEbiography}

\vskip -1\baselineskip
\begin{IEEEbiography}[{\includegraphics[width=1in,height=1.25in,clip,keepaspectratio]{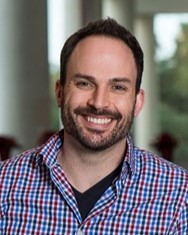}}]{Jacob Robinson} (Senior Member, IEEE) is a Professor in Electrical \& Computer Engineering and Bioengineering at Rice University, where his group develops miniature technologies to manipulate and monitor neural circuit activity. He received a B.S. in Physics from UCLA, a Ph.D. in Applied Physics from Cornell University, and completed Postdoctoral training in the Chemistry Department at Harvard. He previously served as the co-chair of the IEEE Brain Initiative and a core member of the IEEE Brain Neuroethics working group and is currently a member of the IEEE EMBS AdCom. In addition to his academic work, Dr. Robinson is the co-founder and CEO of Motif Neurotech, which is developing minimally invasive bioelectronics to treat mental health disorders.
\end{IEEEbiography}

\vskip -1\baselineskip
\begin{IEEEbiography}
[{\includegraphics[width=1in,height=1.25in,clip,keepaspectratio]{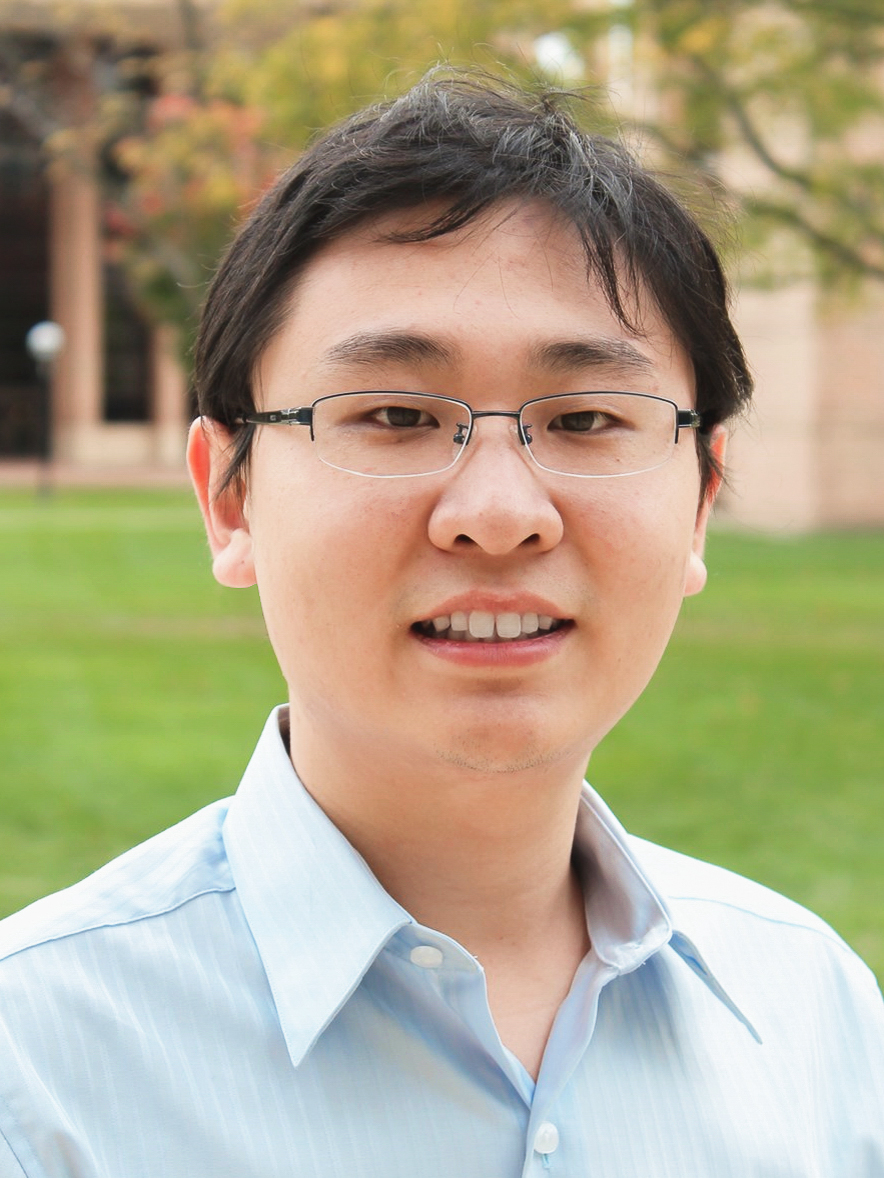}}] {Kaiyuan Yang} (Senior Member, IEEE) is an Associate Professor of Electrical and Computer Engineering at Rice University, USA, where he leads the Secure and Intelligent Micro-Systems (SIMS) lab. He received a B.S. degree in Electronic Engineering from Tsinghua University, China, in 2012, and a Ph.D. degree in Electrical Engineering from the University of Michigan - Ann Arbor, MI, in 2017. His research focuses on low-power integrated circuits and system design for bioelectronics, hardware security, and mixed-signal/in-memory computing. 

Dr. Yang is a recipient of NSF CAREER Award, IEEE SSCS New Frontier Award, SSCS Predoctoral Achievement Award, and best paper awards from premier conferences in multiple fields, including 2024 Annual International Conference of the IEEE Engineering in Medicine and Biology Society (EMBC), 2022 ACM Annual International Conference on Mobile Computing and Networking (MobiCom), 2021 IEEE Custom Integrated Circuit Conference (CICC), 2016 IEEE International Symposium on Security and Privacy (Oakland), and 2015 IEEE International Symposium on Circuits and Systems (ISCAS). His research was also selected as the research highlights of Communications of ACM and ACM GetMobile magazines, and IEEE Top Picks in Hardware and Embedded Security. He currently serves as an associate editor of IEEE Transactions on VLSI Systems (TVLSI) and a program committee member of ISSCC, CICC, ISCA, and MICRO conferences. 
\end{IEEEbiography}

\vspace{11pt}

\vspace{11pt}

\vfill

\end{document}